\documentclass[review]{elsarticle}

\usepackage{multirow}
\usepackage{lineno,hyperref}
\usepackage{color}
\usepackage[utf8]{inputenc}
\usepackage{blindtext, graphicx,nicefrac}
\usepackage{algorithm}
\usepackage{algorithmic}
\usepackage{amssymb}
\usepackage{url}
\usepackage{algorithmic}
\usepackage{soul}
\usepackage{amsmath}
\usepackage{relsize}

\usepackage[table,xcdraw]{xcolor}
\usepackage{csquotes}
\usepackage{subfig}
\usepackage{flushend}
\usepackage{csquotes}
\usepackage{booktabs}
\usepackage{comment}
\usepackage{mathrsfs}
\usepackage{gensymb}
\usepackage{enumerate}
\usepackage{tikz}
\usepackage{lineno}
\usepackage{xcolor}
\usepackage{tablefootnote}

\journal{Engineering Applications of Artificial Intelligence}

\biboptions{authoryear}

\begin{document}
\sloppy
\begin{frontmatter}

\title{RISE Controller Tuning and System Identification Through Machine Learning for Human Lower Limb Rehabilitation via Neuromuscular Electrical Stimulation}

 \author[add1,add2]{H\'eber H. Arcolezi\corref{cor1}}
 \ead{heber.hwang\_arcolezi@univ-fcomte.fr}

 \author[add3]{Willian R. B. M. Nunes}
 \ead{willianr@utfpr.edu.br}
 
 \author[add2]{Rafael A. de Araujo}
 \ead{rafael.araujo@unesp.br}

 \author[add1]{Selene Cerna}
 \ead{selene\_leya.cerna\_nahuis@univ-fcomte.fr}

 \author[add2]{Marcelo A. A. Sanches}
 \ead{marcelo.sanches@unesp.br}

\author[add2]{Marcelo C. M. Teixeira}
\ead{marcelo.minhoto@unesp.br}

\author[add2]{Aparecido A. de Carvalho}
\ead{aa.carvalho@unesp.br}

\cortext[cor1]{Corresponding author}

\address[add1]{Femto-ST Institute, Univ. Bourgogne Franche-Comt\'e, UBFC, CNRS, Belfort, 90000, France}
\address[add2]{Department of Electrical Engineering, S\~ao Paulo State University, UNESP, Ilha Solteira, 15385-000, Brazil}
\address[add3]{Department of Electrical Engineering, Federal University of Technology - Paran\'a, UTFPR, Apucarana, 86812-460, Brazil}

\begin{abstract}

\noindent Neuromuscular electrical stimulation (NMES) has been effectively applied in many rehabilitation treatments of individuals with spinal cord injury (SCI). In this context, we introduce a novel, robust, and intelligent control-based methodology to closed-loop NMES systems. Our approach utilizes a robust control law to guarantee system stability and machine learning tools to optimize both the controller parameters and system identification. Regarding the latter, we introduce the use of past rehabilitation data to build more realistic data-driven identified models. Furthermore, we apply the proposed methodology for the rehabilitation of lower limbs using a control technique named the robust integral of the sign of the error (RISE), an offline improved genetic algorithm optimizer, and neural network models. Although in the literature, the RISE controller presented good results on healthy subjects, without any fine-tuning method, a trial and error approach would quickly lead to muscle fatigue for individuals with SCI. In this paper, for the first time, the RISE controller is evaluated with two paraplegic subjects in one stimulation session and with seven healthy individuals in at least two and at most five sessions. The results showed that the proposed approach provided a better control performance than empirical tuning, which can avoid premature fatigue on NMES-based clinical procedures.

\end{abstract}


\end{frontmatter}

\section{Introduction}
Neuromuscular electrical stimulation (NMES) and functional electrical stimulation (FES) have been effectively applied in many rehabilitation treatments for people with spinal cord injury (SCI) in the past years. Damages in the spinal cord may be engendered by traumatic causes such as road accidents, sports injuries, and violence, or nontraumatic ones such as diseases and tumors. Spinal cord injury is commonly a permanent cause, which can generate issues such as loss of bodily perception, difficulties related to sexual functions, partial or total paralysis, and severe pain (\cite{Ho2014,LynchFES,Popovi2014,Wagner2018}). However, the main consequences depend on several factors, such as the patient's personal condition, the level of the lesion and its damages, the availability of time and resources, and socioeconomic factors. For instance, in low-income countries, SCI normally leads to death, whereas in high-income \textcolor{black}{countries}, people with SCI enjoy a better and more productive life (\cite{who_sci}).

The application of NMES/FES for SCI rehabilitation is one of the most frequently used methods (\cite{marquez2020functional,kapadia2020functional}). It provides many health and social benefits to patients; for example, it helps to preserve and recover muscle strength and prevent flaccidity and hypotrophy, which are evidence of muscle inactivity; it also offers higher expectation and quality of life, and facilitates social reinsertion (\cite{Peckham2005,marquez2020functional,LynchFES}). Moreover, NMES/FES are techniques based on the use of equipment that generates electrical signals for muscle stimulation at the motor level. More specifically, the aim is generating a muscle contraction via electrodes placed superficially or intramuscularly. The electrical stimulation consists of applying a pulsed current or voltage signal that can depolarize neurons above the activation threshold. The amplitude, pulse width (PW), frequency, and shape of the pulse determine which neurons are recruited. Muscle control can be realized by amplitude, PW, or frequency modulation (\cite{Lynch2012,Popovi2014}). 

Even though there are several investigations on the closed-loop control of NMES/FES systems for lower limb rehabilitation (\textit{cf.}~\cite{Ferrarin2001,Previdi2003,Jezernik2004,Cheng2016,Mohammed2012,Wu2017,BenHmed2017,Sharma2012,Gaino2017,Santos2015,nunes2018,Teodoro2020,Mller2017,Page2020,Gaino2020} and the references within), these systems are hardly put into production. Alternatively, there exist commercial stimulators normally available on open-loop designs and with pre-programmed electrical stimulation, which are not adequate to deal with the nonlinear and time-varying nature of muscles (\cite{Page2020,Lynch2012}). 
Hence, given the numerous challenges in the design of automatic stimulation strategies, further investigation is needed in this field. For example, control strategies are needed to compensate for modeling errors on the plant, system faults, individual's muscles behavior, and inter/intra-subject variability in muscle properties (\cite{Sharma2009,Sharma2012,Yu2013,Yu2015}). The variability in muscle properties leads to the difficulty of predicting the exact contraction force exerted by the muscle, which results in unknown mapping between the stimulation parameters and the muscle force. 


In this sense, the design and evaluation of the robust integral of the sign of the error (RISE) control (\cite{Xian2003, Xian2004}) for tracking the nonlinear dynamics of electrically stimulated lower limbs are presented in this paper. Despite several control laws investigated in the literature, this study considers RISE control law by some fundamental characteristics, such as the consideration of unmodeled disturbances and uncertainties in the plant. Nevertheless, adjusting the controller parameters is the main component to guarantee high-quality control performance; that is, the method can only guarantee good responses (semi-global asymptotic stability), appropriately selecting the gain constants.

\cite{Stegath2007, Stegath2008} and~\cite{Sharma2009} are pioneers authors on RISE controller development for the lower limb tracking control. Afterward,~\cite{Sharma2012} presented an improvement of RISE control method for the same application using a feedforward neural network (NN) term.~\cite{Downey2013} and~\cite{Downey2015_b} developed an RISE controller for the asynchronous stimulation to the lower limb.~\cite{Kawai2014} simulated the tracking control performance of an RISE-based controller to model the co-contraction control of the human lower limb.~\cite{Kushima2015} modeled an FES knee bending and stretching system, and developed an RISE-based controller to stimulate the quadriceps and hamstring muscle groups. In the similar context of NMES, but for upper limbs,~\cite{Lew2016} implemented RISE controller for the rehabilitation of post-stroke individuals.

Even though previous investigations for this problem with RISE controller presented good results without any fine-tuning method, the motivation of this paper is the absence of clever algorithms to properly select the gain constants of RISE controller. In the aforementioned studies, the authors did not show the controller tuning method or empirical approach (pretrial tests) for defining gain parameters before the real experiments are conducted. In addition, experiment validations were made only on healthy individuals; however, the muscles of people with SCI are not as strong as healthy muscles (\cite{Mohammed2012,Lynch2012}).

More specifically, \cite{Stegath2008}, \cite{Sharma2009}, and \cite{Downey2015_b} present four inequalities to gain constants, which are sufficient conditions to guarantee semi-global asymptotic stability for an uncertain nonlinear muscle model. There are infinite combinations of gains in $\mathbb{R}^+$ that satisfy these inequalities; yet, as presented in the aforementioned works, a \enquote{trial and error} methodology might be feasible to set gain constants to the controller for healthy subjects. However, this procedure must be reconsidered when treating people with SCI to avoid some common problems. For instance, for SCI rehabilitation via NMES/FES, there might exist rapid muscle fatigue, muscle tremors due to incomplete tetanus, and harsh muscle spasms (\cite{Ho2014,Lynch2012,Popovi2014,Peckham2005}).

Therefore, to overcome the aforementioned problems, this paper proposes a novel robust and intelligent control-based methodology for NMES/FES systems. More precisely, we aim to overcome the empirical tuning technique for clinical procedures using RISE controller, as observed in the literature. Moreover, this study proposes to extend the analysis of RISE controller to individuals with SCI that do not present ideal conditions as healthy individuals. The proposed methodology includes an identification step based on machine learning (ML) black-box models with the novelty of using past identification and control data for each patient, a robust control law (e.g., RISE technique) to guarantee the semi-global asymptotic stability, and an ML-based offline controller optimizer.

In \cite{arcolezi2019}, our group proposed an offline improved genetic algorithm (IGA) optimizer to RISE controller. Simulations were performed using a nonlinear mathematical model of the knee joint for three paraplegics and one healthy individual. In this study, our proposed methodology is implemented and evaluated with seven healthy and two paraplegic individuals using RISE control law, the aforementioned IGA optimizer, and NN black-box models. 

The first hypothesis in this paper is that using an empirical approach to clinical procedures would present a large number of poor performances, while a more adequate tuning with a more representative identified model can provide better tracking control of the lower limb. \textcolor{black}{That is, we assume no background knowledge with the RISE controller for clinicians intending to design and apply it to real-life scenarios.} The second hypothesis is that by using past rehabilitation data for identifying an individual, this model will improve the description of the relationship between the angular position and the delivered electrical stimulation, whereby fatigue and other problems as tremors are already implicit in the data.

The remaining sections of this paper are organized as follows: Section~\ref{sec:theoretical} presents the theoretical background; Section~\ref{sec:mat_metho} introduces the proposed control-based methodology and the materials and methods used in the experiments; Section~\ref{sec:resul} presents the results and its analysis; and finally, Section~\ref{sec:conc} provides the conclusions of this paper and future works.

\section{Theoretical Background}
\label{sec:theoretical}

In this section, we briefly present the musculoskeletal dynamics about the knee joint (Subsection~\ref{sub:Dynamic}) and RISE control method (Subsection~\ref{sub:RISE}). We summarize the IGA for the optimization procedure (Subsection~\ref{sub:IGA}), and we discuss nonlinear system identification via NN models (Subsection~\ref{sub:NNs}).

\subsection{System dynamics}\label{sub:Dynamic}
The musculoskeletal dynamics based on electrical stimulation is given as (\cite{Ferrarin2001,Sharma2009})
\begin{equation}
     J\ddot{\theta}(t) =\Lambda_{g}(\theta(t))+ \Lambda_{e}(\theta (t)) + \Lambda_{v}(\dot{\theta}(t)) + \Lambda_{d}(t) +\Lambda_{es}(t),
     \label{eq:dynamic}
\end{equation}
\noindent where $J \in \mathbb{R}$ is the unknown inertia of the combined shank and foot; $\theta(t),\; \dot{\theta}(t),\; \ddot{\theta}(t) \in \mathbb{R}$ is the angular position, velocity and acceleration, respectively.

The gravitational component $\Lambda_{g}(\theta(t)) \in \mathbb{R}$ is expressed as
 \begin{equation}
    \Lambda_{g}(\theta(t))=-mgl\sin(\theta(t)),
    \label{eq:lambda_g}
 \end{equation}
 \noindent where $m \in \mathbb{R}$ denotes the unknown combined mass of the shank and foot; $l \in \mathbb{R}$ is the unknown length between the knee-joint and center of mass of the shank and foot; and $g \in \mathbb{R}$ is the gravitational acceleration.
 
The elastic effects due to joint stiffness $\Lambda_{e}(\theta(t)) \in \mathbb{R}$ can be modeled as
 \begin{equation}
     \Lambda_{e}(\theta(t))=-\left(\psi_{1}\theta(t)-\psi_{1}\psi_{3}\right)\left(e^{-\psi_{2}\theta(t)}\right),
     \label{eq:lambda_e}
 \end{equation}
 \noindent where $\psi_{1},\; \psi_{2},\; \psi_{3} \in \mathbb{R}$ are unknown positive coefficients.
 
The viscous effects due to damping $\Lambda_{v}(\dot{\theta}(t)) \in \mathbb{R}$ is defined as
 \begin{equation}
     \Lambda_{v}(\dot{\theta}(t))=-\kappa_{1} \tanh(-\kappa_{2} \dot{\theta}(t))+\kappa_{3}\dot{\theta}(t),
     \label{eq:lambda_v}
 \end{equation}
 \noindent where $\kappa_{1}, \; \kappa_{2},\; \kappa_{3} \in \mathbb{R}$ are unknown positive constants.
 
The torque produced at the knee joint by the electrical stimulation $\Lambda_{es}(t) \in \mathbb{R}$ is related to the positive moment $\varsigma(\theta(t)) \in \mathbb{R}$ from the extension and flexion of the leg, the unknown nonlinear function $\nu(\theta, \dot{\theta}) \in \mathbb{R}$ corresponding to muscle tendon force, and the electrical potential $u(t) \in \mathbb{R}$ applied to the quadriceps muscle:
\begin{equation}
    \Lambda_{es}(t)=\varsigma(\theta(t))\nu(\theta(t), \dot{\theta}(t))u(t).
    \label{eq:lambda_es}
\end{equation}

Finally, $\Lambda_{d}(t) \in \mathbb{R}$ is the unmodeled bounded disturbances (e.g., fatigue, spasms, tremor, and delay).

\subsection{RISE-based control}\label{sub:RISE}

RISE control method proposed by~\cite{Xian2003,Xian2004} utilizes a continuous and high gain control signal, which guarantees semi-global asymptotic stability considering bounded smooth external disturbances and bounded modeling uncertainties. The use of the integral of the sign of the error in RISE technique minimizes the commonly chattering problem seen in sliding-mode controllers. To achieve the stated control objective, i.e., to enable the lower limb to track a desired angular trajectory despite external disturbances and modeling uncertainties, a position tracking error denoted by $e_1(t)\in\mathbb{R}$, is defined as

\begin{equation}
  e_1(t)=\theta_d(t)-\theta(t)\textrm{,} 
  \label{eq:e1}
\end{equation}

\noindent where $\theta_d(t)$ is the angular trajectory to be tracked with the premise of having bounded continuous-time derivatives, and $\theta(t)$ is the real angular position. Furthermore, to facilitate the control design, filtered tracking errors $e_2(t)\in\mathbb{R}$ and $r(t)\in\mathbb{R}$ are defined as

\begin{equation}
  e_2(t)=\dot{e}_1(t)+\alpha_1e_1(t)\textrm{,}
  \label{eq:e2}
\end{equation}

\begin{equation}
  r(t)=\dot{e}_2(t)+\alpha_2e_2(t)\textrm{,} 
  \label{eq:r}
\end{equation}

\noindent where $\alpha_1,\alpha_2\in\mathbb{R}$ are positive and selectable control gains. 

\textcolor{black}{Multiplying \eqref{eq:r} by $J$, and considering \eqref{eq:dynamic}-\eqref{eq:e2}, $\dot{e}_2=\ddot{\theta}_{d}(t)+\alpha_1\dot{e}_1-\ddot{\theta}(t)$, one obtains
\begin{equation}
    Jr=\Upsilon(\dot{\theta}_d,\dot{\theta}, \theta, \dot{e}_1,e_2)-\Psi(\dot{\theta},\theta) u-\Lambda_d\textrm{,}
    \label{eq:Jr}
\end{equation}
\noindent where $\Upsilon(\dot{\theta}_d(t),\dot{\theta}(t), \dot{e}_1(t),e_2(t)) \in \mathbb{R}$ defined as
\begin{equation*}
    \Upsilon(\dot{\theta}_d,\dot{\theta},\theta, \dot{e}_1,e_2)=\ddot{\theta}_d+\alpha_1\dot{e}_1+\alpha_2e_2-\Lambda_g(\theta)-\Lambda_e(\theta)-\Lambda_v(\dot{\theta})\textrm{,}
\end{equation*}
\noindent and $\Psi(\theta, \dot{\theta}) \in \mathbb{R}$ a function monotonic and bounded, expressed as 
\begin{equation*}
    \Psi(\dot{\theta},\theta)=\varsigma(\theta)\nu(\theta, \dot{\theta})\textrm{.}
\end{equation*}}

\textcolor{black}{For stability analysis, from \eqref{eq:Jr} can be determined the open-loop error system
\begin{equation*}
    \mathcal{J}_{\Psi}r=\mathcal{Y}_{\Psi}-u-\mathcal{L}_{\Psi}\textrm{,}
\end{equation*}
\noindent where $\mathcal{J}_{\Psi}=\Psi^{-1}J$, $\mathcal{Y}_{\Psi}=\Psi^{-1}\Upsilon$, and $\mathcal{L}_{\Psi}=\Psi^{-1}\Lambda_d$, and consequently one obtains
\begin{equation*}
    \mathcal{J}_{\Psi}\dot{r}=-\dot{u}-e_2+\tilde{\mathcal{W}}+\mathcal{W}_d\textrm{,}
\end{equation*}
\noindent where $\tilde{\mathcal{W}}=\mathcal{W}-\mathcal{W}_d$, $\tilde{\mathcal{W}}(e_1,e_2,r,t) \in \mathbb{R}$, $\mathcal{W} \in \mathbb{R}$ corresponds to the term
\begin{equation*}
    \mathcal{W}=-\frac{1}{2}\dot{\mathcal{J}}_{\Psi}r+\dot{\mathcal{Y}}_{\Psi}-\dot{\mathcal{L}}_{\Psi}+e_2\textrm{,}
\end{equation*}
\noindent \textcolor{black}{and $\mathcal{W}_d \in \mathbb{R}$ expressed as}
\begin{equation*}
    \mathcal{W}_d=\dot{\mathcal{J}}_{\Psi}\ddot{\theta}_d+\mathcal{J}_{\Psi}\dddot{\theta}_d-\dot{\Lambda}_e-\dot{\Lambda}_g-\dot{\Lambda}_v-\dot{\Lambda}_d\textrm{.}
\end{equation*}}

\textcolor{black}{Based on the mean value theorem applied to upper bound $\left \| \tilde{  \mathcal{W}} \right \| \leq \varsigma \left( \left\| \zeta \right\| \right)\left\| \zeta \right\|$, where $\zeta \in \mathbb{R}^3,\; \zeta=\left[r^T  \; e_1^T  \;  e_2^T\right]^T$, $\varsigma\left( \left\|\zeta \right\| \right) \in \mathbb{R}$ is a positive globally invertible nondecreasing function, and considering that $\theta_d$, and its derivatives $\theta_d^{(k)} \in \mathcal{L}_{\infty}, \forall k \in \mathbb{I}=\left\{1,\; 2,\; 3,\; 4 \right\}$, the following constraints can be established $\left \| \mathcal{W}_d \right \| \leq \mathcal{E}_{\mathcal{W}_d}$, $\left \| \dot{\mathcal{W}}_d \right \| \leq \mathcal{E}_{\dot{\mathcal{W}}_d}$, such as $\mathcal{E}_{\mathcal{W}_d}$, $\mathcal{E}_{\dot{\mathcal{W}}_d} \in \mathbb{R}$ are positive constants (\cite{utkin2013sliding}).}

\textcolor{black}{Note that the system error equations obtained to nonlinear dynamic model are similar to other studies with the RISE controller in (\cite{Sharma2009,Stegath2008,Patre2008,Makkar2007,Xian2004,Xian2003}). Based on the open-loop error system, the control input $u(t) \in \mathbb{R}$, is designed as
\begin{equation}
u(t)=(k_s+1)e_2(t)-(k_s+1)e_2(0)+\int_{0}^{t}[(k_s+1)\alpha_2e_2(\tau)+\beta sgn(e_2(\tau))]d\tau\textrm{,}
\label{eq:u_RISE}
\end{equation}
\noindent where $k_s,\beta\in\mathbb{R}$ also represents positive and adjustable control gains, $u(t)$ is the control signal, and $sgn(\cdot)$ is the known signum function.}

\textcolor{black}{The RISE controller, given in \eqref{eq:u_RISE}, ensures that all system signals are bounded under closed-loop operation and the position tracking error is regulated in sense that 
\begin{equation*}
    \lim_{t \rightarrow \infty}\left \|e_1(t) \right|\rightarrow 0 \textrm{,}
\end{equation*}
\noindent yields semi-global asymptotic stability provided the control gain $k_s$ sufficiently large, and $\beta$ satisfying the following sufficient condition
\begin{equation}
    \beta > \mathcal{E}_{\mathcal{W}_d}+\frac{1}{\alpha_2}\mathcal{E}_{\dot{\mathcal{W}}_d} \textrm{,}
    \label{eq:betaRISE}
\end{equation}
\noindent where $\mathcal{E}_{\mathcal{W}_d},\;\mathcal{E}_{\dot{\mathcal{W}}_d} \in \mathbb{R}$ are known positive constants. More details about the stability analysis of the RISE method can be found in (\cite{Patre2008, Makkar2007, Xian2004}).}



The ideal first derivative of the error $H(s)=\dfrac{Y(s)}{U(s)}=\dfrac{sE(s)}{E(s)}=s$ is an improper function, that is, $H(s)=\dfrac{\sum_{j=0}^{m} b_j s^j}{\sum_{i=0}^{n} a_i s^i}$, $a_i,\;b_j \in \mathbb{R}$, $\forall i=1,\;2,\; \cdots,\; n$, $\forall j=1,\;2,\; \cdots,\; m$, $m>n$, $\left | H(\infty)  \right |=\infty$. The unfeasibility of practical implementation using the ideal derivative is solved by a filtered derivative (\cite{AbuKhadra2016}). Thus, the filtered tracking error is calculated by
    \begin{equation}
        H(s)=\frac{Y(s)}{U(s)}=\frac{s}{\tau s+1},
        \label{eq:filtder}
    \end{equation}
\noindent where $\tau$ is the time constant between the signal and its derivative. Note that~\eqref{eq:filtder} is a low pass filter (LPF) that attenuates high-frequency noises.

\subsection{Improved genetic algorithm}\label{sub:IGA}

The IGA was introduced in \cite{arcolezi2019} to optimize the gains parameters of RISE controller for a representative model of an individual. This algorithm is summarized in this paper. \textcolor{black}{First, there is a pre-processing stage for bounding the gain limits to efficiently initiate (i.e., the random initial population within the constraints of stability) and maintain the search (i.e., genetic operators such as recombination, mutation, and replacement operator).} Second, a simple fast genetic algorithm (FGA) is used in the construction phase to generate a good initial population. Thereafter, a complete genetic algorithm (CGA) is applied to improve the quality of this population and hence achieve a global (or local) minimum. 

\begin{figure}[!h]
	\centering
	\scriptsize
	\includegraphics[width=0.5\linewidth]{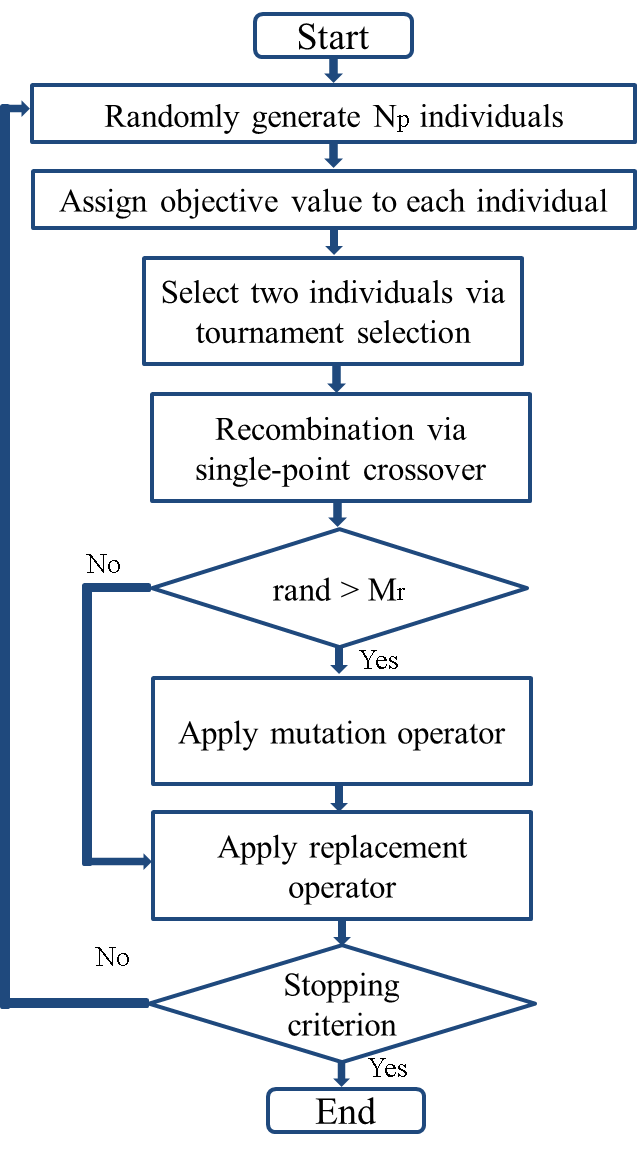}
	\caption{Flow chart of the fast genetic algorithm.}
	\label{fig:iga}
\end{figure}

Figure~\ref{fig:iga} describes the FGA with a flow chart. In the chart, $N_p$ is the size of the initial population (small), $M_r$ is the mutation rate, and the stopping criterion is the number of generations $N_g$. More specifically, $N_g$ represents the size of the real initial population (RIP) to initiate the local search phase. The CGA is similar to the FGA, with a more stringent test to the replacement operator. We recommend that readers refer to (\cite{arcolezi2019}) for a more descriptive version of the algorithm. 

\subsection{System identification via neural networks}\label{sub:NNs}

Nonlinear systems identification and modeling have been applied in most areas of science to predict the future behavior of dynamic systems. System identification has been an active field in control theory, and it is an important approach to explore, study, and understand the world by a formal description of events as a model. \textcolor{black}{The use of NNs to identify nonlinear systems has been a prospective direction since previous research presented in (\cite{hornik1989multilayer,CHEN1990,Narendra1990,Chu1990}), for example.} In the following, the use of NNs for the identification of discrete dynamic system is briefly described.

The construction of black-box models is essentially based on the quality of measured data about the system. The fundamental concept of this approach is to model the direct input-output relationship, i.e., identifying and modeling just with data, in which the main objective is to find the weights and other coefficients (known as hyperparameters) of the NN. Moreover, NNs are based on a collection of inter-connected units named neurons. These neurons are structured into three or more layers, input, hidden(s), and output. Neural networks are in the core of deep learning (several neurons and hidden layers) and have become a progressively popular research topic. Generally, NNs can be divided into two large classes: feedforward and recurrent NNs.

Fundamentally, an operator \(F\) from an input space \(\mathbb{U}\) to an output space \(\mathbb{Y}\) expresses the model of the system to be identified, where the goal is to find a function \(\hat{F}\) that approximates \(F\) to a specific requirement. By the Stone-Weierstrass theorem, there exists a continuous and bounded function \(F\), that can be uniformly approximated as closely as desired by a polynomial function \(\hat{F}\). Furthermore, according to the universal approximation theorem, there exists a combination of hyperparameters of an NN that allows it to identify and learn any continuous nonlinear function defined on a closed interval \textcolor{black}{(\cite{hornik1989multilayer}).} 

Consider a single-input and single-output discrete system structure with only the input and output data available:

\begin{equation}  y(k) = F[y(k-1),...,y(k-n);u(k-1),...,u(k-m)]\textrm{,}  \label{eq:siso} \end{equation}

\noindent where \(F(\cdot)\) is an unknown nonlinear difference equation that represents the plant dynamics; \(u\) and \(y\) are measurable scalar input and output, respectively; and \(m\) and \(n\) are the maximum lags for the system output and input; that is, they are the last values of the input and output respectively. In short, the next value of the dependent output signal \(y(k)\) is regressed on previous values of the output and input signals.

The identification for the discrete-time system in~\eqref{eq:siso} can be performed by the following two major types of identification structures presented in the literature: the parallel and the series-parallel identification model (\cite{Narendra1990}). The first structure depends on past inputs of the system and the outputs of the NN model. The second structure uses both past inputs and system's outputs. Mathematically, these models are respectively described as

\begin{equation}  \hat{y}(k) = \hat{F}[\hat{y}(k-1),...,\hat{y}(k-n);u(k-1),...,u(k-m)]\textrm{,}  \label{eq:parallel} \end{equation}
\begin{equation} \hat{y}(k) = \hat{F}[y(k-1),...,y(k-n);u(k-1),...,u(k-m)]\textrm{,}  \label{eq:series_parallel} \end{equation}

\noindent where \(\hat{y}\) is the model output; \(y\) is the real system output; \(\hat{F}\) is the model structure; and \(m\) and \(n\) are the regression orders for the input and output, respectively. These last two parameters are chosen before the identification process, where \(n\) is the output memory to indicate how many past steps of output will be used in the system identification, and \(m\) refers to the time-step of input values and it is the longest memory that a model can store. In this paper, we used a feedforward NN (multilayer perceptron - MLP) to approximate the nonlinear mapping function \(F(\cdot)\) in~\eqref{eq:siso} using the series parallel structure in~\eqref{eq:series_parallel}.

\section{Materials and Methods}
In this section, we first present an overview of our proposed methodology (Subsection~\ref{sub:prop_metho}). Next, we provide information on the volunteering participants (Subsection~\ref{sub:individuals}) and on the instrumentation used for real experiments (Subsection~\ref{sub:instru}). Lastly, we describe how we applied the proposed methodology in this study for both, data acquisition and experimental procedures (Subsection~\ref{sub:experiments}).

\label{sec:mat_metho}

\subsection{Proposed methodology}\label{sub:prop_metho}

\begin{figure}[!htb]
	\centering
	\scriptsize
	\includegraphics[width=1\linewidth]{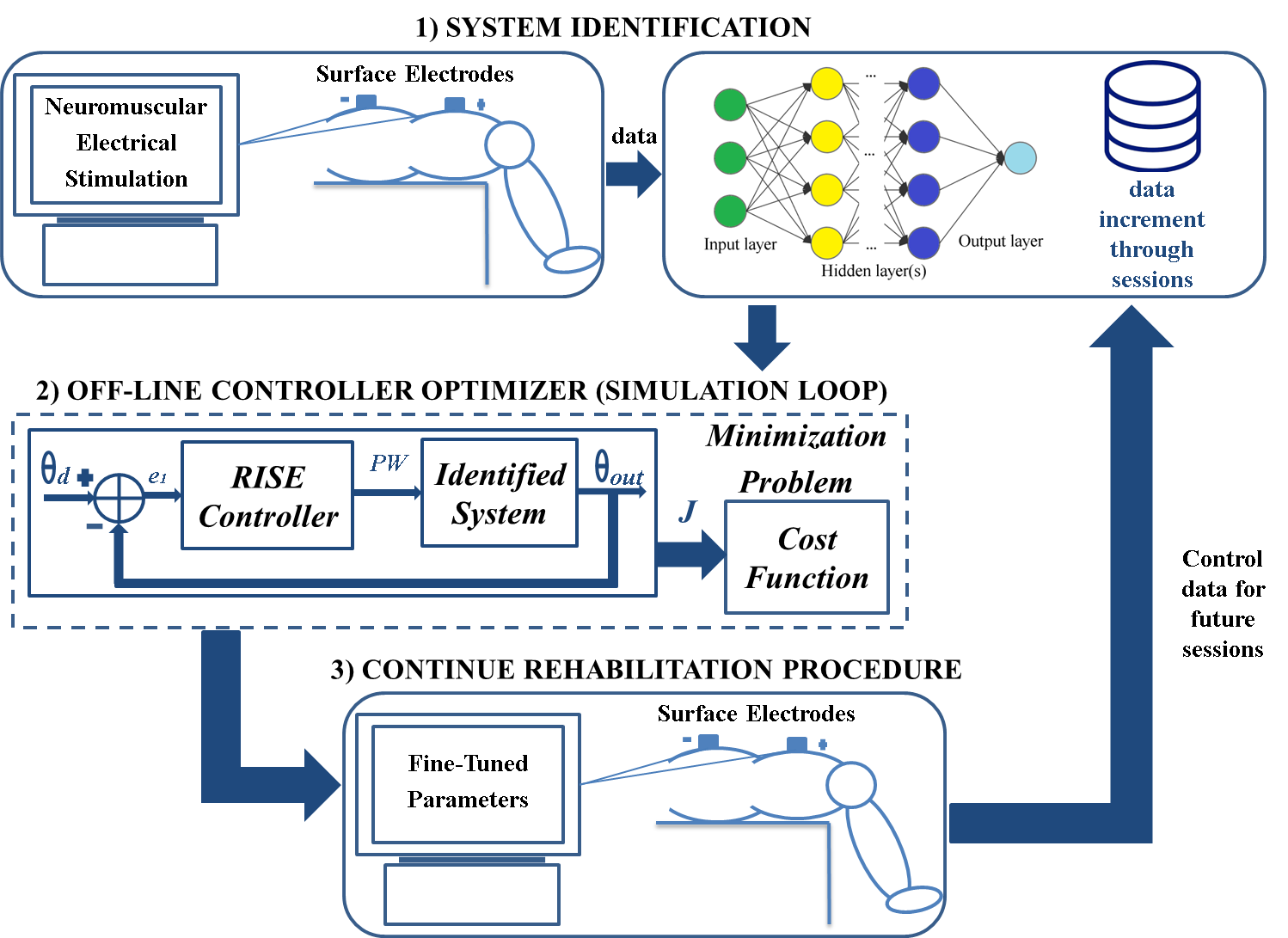}
	\caption{The proposed robust and intelligent control-based methodology.}
	\label{fig:comp_metho}
\end{figure}

Fig.~\ref{fig:comp_metho} illustrates an overview of the proposed methodology. In the first session of a new patient (no previous data), a stimulation test is performed to acquire information on the relationship between delivered electrical stimulation and the achieved angular position. The acquired data are appropriately treated to pass through an identification step via NN black-box models. Once this relationship is efficiently mapped as a model, a simulation process is initiated using clever algorithms. The aim is to minimize a well-defined objective function to adequately set-up the gains of RISE controller for the patient. Therefore, to finalize the first session, the rehabilitation procedure is retaken with fine-tuned gains for a better control-stimulation session. This could prevent premature fatigue and other unwanted factors that would be present for people with SCI by not choosing an appropriate gain combination. 

In future sessions, all data (system identification and control evaluation) from previous rehabilitation sessions are used for training an NN model in an offline scheme. That is, before each (next) session, all data from a patient are combined to a single dataset and used to map the relationship between angular position and electrical stimulation. Thus, the same optimization process using the trained model provides fine-tuned gain parameters to be afterward applied to the rehabilitation procedure. The gains of the controller are found using only past rehabilitation data, which is motivated by the belief that preliminary electrical stimulation could lead to quick muscle fatigue during the real clinical procedure. Moreover, between stimulation sessions, there exist factors such as fatigue, hydration, evolution/gain of strength, rest, and therapeutic sessions, which might influence one's response to NMES/FES and make the control-stimulation inefficiently.

The use of NNs is motivated by the advantages of these methods for the nonlinear system identification problem and by the high power for computation and storage of data encountered nowadays. Regarding the identification step, the novelty of the proposed methodology is the use of past rehabilitation data. The primary purpose is to build up a dataset for each patient, where the number of data will increase during rehabilitation sessions, and the identified model will improve with more data and details about the nonlinear muscular behavior. As highlighted in the literature, muscular behavior is susceptible to parametric variation between one day and another, for instance, the evolution and gain of strength due to previous rehabilitation sessions. 

Moreover, one of the primary advantages of performing simulations for an NMES-based knee extension is the liberty of studying this problem from different perspectives and divergent levels of abstraction with the acquired data. While the application of NMES to humans presents limitations due to muscle fatigue, which restricts the number of experiments, simulation provides numerous executions to better study the feasibility and practicality of the designed system. Moreover, simulation supplies continuous feedback to continuously improve the system (\cite{Jezernik2004}).

\subsection{Analyzed individuals}\label{sub:individuals}

The study with volunteers was authorized through a research ethics committee involving human beings (CAAE: 79219317.2.1001.5402) at São Paulo State University (UNESP). Written informed consent was obtained from all participants before their participation. In this study, seven healthy individuals (male, aged 22-28) labeled as H1-H7 and two male individuals with SCI, labeled as P1 and P2, participated in the experiments. Table~\ref{tbl_parameters_unesp} presents information on the two SCI individuals, including age, injury data, and ASIA (American Spinal Injury Association) Impairment Scale (AIS).

\begin{table}[!htb]
    \centering
    \caption{Specific data on individuals with SCI.}
    \label{tbl_parameters_unesp}
    \begin{tabular}{c c c c c}
    \hline
    \textbf{Individual} &\textbf{Age (years)} &\textbf{Injury level}&\textbf{Injury time} & \textbf{AIS}\\
    \hline
	 P1 &32  &L4, L5 & 9 years & B \\
	 P2 &43  &C5, C6 & 17 years & C\\
    \hline
    \end{tabular}
\end{table}  

\subsection{Instrumentation}
\label{sub:instru}

\begin{figure}[!h]
	\centering
	\includegraphics[width=1\linewidth]{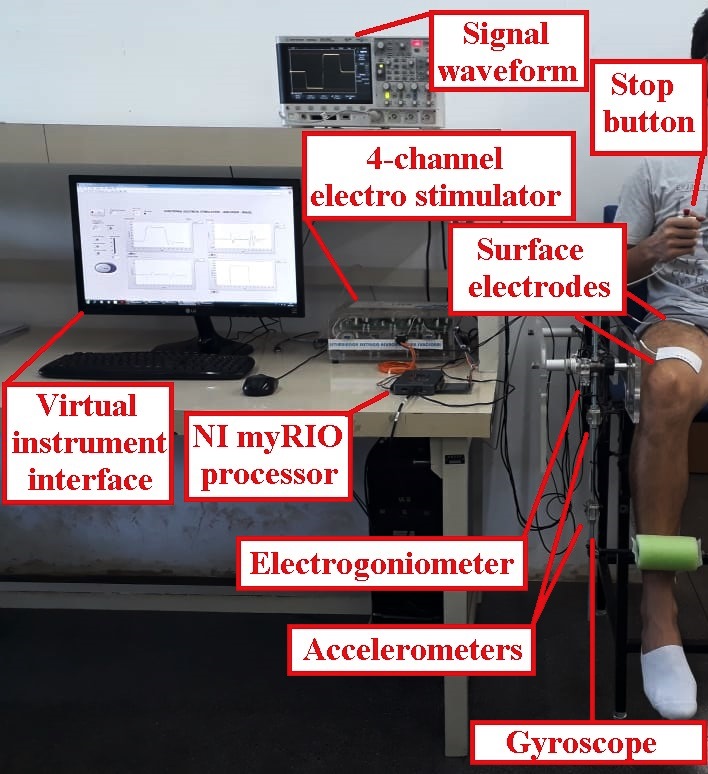}
	\caption{Test platform for electrical stimulation experiments.}
	\label{fig:test_platform}
\end{figure}

Fig.~\ref{fig:test_platform} illustrates the test platform used for conducting the experiments at the Instrumentation and Biomedical Engineering Laboratory (\enquote{Laboratório de Instrumentação e Engenharia Biomédica - LIEB}) at UNESP - Ilha Solteira. The platform was composed of an NI (National Instruments\textsuperscript{\tiny\textregistered}, USA) myRIO controller to operate in real time; a current-based neuromuscular electrical stimulator; an instrumented chair composed of an electrogoniometer NIP 01517.0001 (Lynx\textsuperscript{\tiny\textregistered}, São Paulo, Brazil), a gyroscope LPR510AL (ST Microelectronics\textsuperscript{\tiny\textregistered}, Switzerland), two triaxial accelerometers MMA7341 (Freescale\textsuperscript{\tiny\textregistered}, USA); and two user interfaces developed in LabVIEW\textsuperscript{\tiny\textregistered}, one for identification and the other for controlling.

The neuromuscular electrical stimulator delivers rectangular, biphasic, symmetrical pulses to the individual's muscle, allowing a control adjustment of the PW in a range of $0-400\mu s$. We controlled the stimulation intensity by setting the pulse amplitude to the quadriceps and controlling the PW. In this study, we fixed the following parameters: stimulation frequency at 25 Hz (constant frequency train - CFT technique) and pulse amplitude at 80 mA for healthy individuals and 120 mA for paraplegic ones. The difference in pulse amplitude occurred due to insufficient contractions using amplitude below 120 mA for the paraplegic individuals and their respective muscular atrophy conditions. Lastly, we used surface electrodes with rectangular self-adhesive CARCI 50 mm x 90 mm settings.

\subsection{Data acquisition and experimental procedure}
\label{sub:experiments}

The chair backrest and the knee joint position were adjusted to ensure the volunteers' comfort. Each individual had a different knee angular position in the resting condition. The angular position in this condition was measured and taken as an offset during the experimental protocol. A muscle analysis was conducted to determine the motor point and guarantee the proper positioning of the surface electrodes. More precisely, the electrophysiological procedure for identifying the motor point consists of mapping the muscle surface using a stimulation electrode to identify the skin area above the muscle, where the motor threshold is the lowest for a given electrical current; this skin area is the most responsive to electrical stimulation (\cite{Gobbo2014}). After this procedure, the electrodes were properly positioned allowing the neuromuscular electrical stimulation to maximize the effectiveness of the evoked voltage, minimizing the intensity of the injected current and the level of discomfort to the volunteer.

After the motor-point identification, a few open-loop tests were performed by applying a step input during four seconds. It is worth highlighting the definition of the electrical current level of the stimulator, as well as evaluating the PW values for different operating points of the lower limb extension. If the value $\rho_{max}$ tends to the saturation value of the stimulator, the electrical current amplitude must be increased so that the control system adequately compensates for disturbances and uncertainties in the process. Moreover, the $\rho_{min}$ is related to the minimum joint extension value from the resting position. In this study, the tests were performed to obtain $\rho_{max}$ and $\rho_{min}$ corresponding to $\theta_{max}=40\degree$ and $\theta_{min}=10\degree$, respectively. Note that we could adopt other values of lower limb extension, but we consider that it was a suitable value for gait control application (\cite{nunes2018}). Lastly, $\rho_{min}$ and $\rho_{max}$ were also useful to select the initial PW and an upper bound to the control signal, respectively. \textcolor{black}{If $\rho_{max}$ does not approach the saturation value of the stimulator ($400\mu s$), with the consent of each individual, we select an adequate upper bound to the control signal for each stimulation session, aiming to minimize the discomfort level to volunteers.}

During the experiments, healthy individuals were instructed to relax, to not influence the leg motion voluntarily, and allow the stimulation to control it. During electrical stimulation sessions, the individuals could deactivate the stimulation pulses using a stop button under any uncomfortable situation (as shown in Fig.~\ref{fig:test_platform}).

In the following two subsections, the experimental setup is detailed. First, the case of an individual using the proposed methodology for the first time, i.e., without any previous data, is considered (Subsection~\ref{subsec:first_session}). Next, the case of individuals who participate in more than one rehabilitation session is considered (Subsection~\ref{subsec:two_five_sessions}).

\subsubsection{First session}
\label{subsec:first_session}

\textcolor{black}{In the first session of a new patient, a one-minute stimulation test was conducted.} In this stimulation, the experimental system identification procedure was performed by randomly applying PW values belonging to the set of values mapped to each individual. The electrical PW random value was constant for a random time between four and seven seconds. Consequently, a new test has randomness in the domain of the PW of the electrical stimulation as well as in the time of each stimulation. In this work, the power of muscle activation by electrical stimulation in paraplegic individuals was greater. Before the tests were performed, these individuals were not admitted to a rehabilitation research program involving daily electrically stimulated exercise of their lower limbs. Consequently, under high stimulation intensity, there was only partial recruitment of synergistic motor units and there was the co-activation of antagonists (\cite{doucet}). Unfortunately, this is a disadvantage of conventional single-electrode stimulation, whose increased stimulation intensity will lead to increased muscle fatigue (\cite{Laubacher2017,Maffiuletti2010}). To minimize early fatigue in paraplegic individuals (\cite{Gregory2007}), the total test time was reduced to 40 $s$.

The motivation to adopt this methodology is to map a tracking situation and recognize the completely nonlinear and time-varying nature of muscles under long electrical stimulation time. The PW ($\mu s$) and angular position (rad) data were automatically recorded with a sampling period of 20 $ms$, i.e., $Ts=0.02$ ($s$), resulting in datasets with approximately 3000 samples (60 $s$) at most.

Afterward, the identification data were read and manipulated for feeding up a shallow MLP with one hidden layer. In the literature, one hidden layer has been proved to be sufficient to approximate any continuous function on a compact domain (\cite{hornik1989multilayer,Previdi2002}). We tuned the number of neurons via a random search procedure (\cite{Bergstra12randomsearch}), in which a combination of hyperparameters is randomly selected to find the best solution for the built model. This process was only done for individual H1, which was the first volunteer for this study, and it took less than 30 min to find an appropriate architecture to be used for all other individuals. The number of neurons was selected as 250; hyperbolic tangent activation was used in each neuron from the hidden layer, and the output layer was composed of one neuron with linear activation, which gives the estimated output \(\hat{y}(k)\).

We experimented with several $m$ and $n$ values, and the one with the best time-utility trade-off was $m=n=1$. This resulted in datasets containing the last input value \enquote{Pulse\_Width\((k-1)\)} and the last output value \enquote{Angular\_Position \((k-1)\)} as features, and the actual output value \enquote{Angular\_Position \((k)\)} as target. The MLP NN model requires a normal input arranged as \([samples, features]\), where the observations at previous time-steps are inputted as features to the model. \textcolor{black}{In general, the training time of each NN model in the first session did not exceed 5 min as the number of samples was small ($\sim 3,000$ for healthy individuals and $\sim 2,000$ for SCI ones).}

Therefore, using the estimated model, we performed an optimization procedure based on the proposed IGA to find the best gains combination for two reference trajectories. The first trajectory is a sinusoidal wave ranging from $10\degree$ to $40\degree$ and the second trajectory is a $40\degree$ step wave ($30\degree$ for individuals with SCI); the first and second trajectories simulate isotonic and isometric contractions, respectively. A smooth range of motion at $40\degree$ and a small-time period (sine wave) was used to avoid premature fatigue by diminishing muscle effort.

Considering a real-world application of the proposed methodology and by assuming a limited time for a rehabilitation session, we used the following as the initial parameters of the IGA simulations: population size $N_p=8$, mutation rate \(M_r=0.5\), number of generations \(N_g=6\) (size of RIP), and \(k=1\) iteration. The algorithm ran only once providing $N_g$ combination of RISE controller gains. Generally, the running time did not exceed 10 min of execution. 

\textcolor{black}{Notice that the proposed IGA in \cite{arcolezi2019} has a pre-processing step (step 1 of the algorithm), which tries to bound the gain values when applying the genetic operators (crossover, mutation) to the required stability conditions presented in Subsection~\ref{sub:RISE}}. However, there is still a possibility that given an identified model and the optimization procedure that gain values deviate from the required conditions. \textcolor{black}{Yet, as genetic algorithms are population-based, one can compare and select the most appropriate combination of gains for a given individual that satisfies the gain's condition. Before the real experiment, previous simulations of both trajectories were made to visually inspect the system response.}

Lastly, using empirical gains and the ones encountered by the IGA, the controlling procedure was implemented for both trajectories. Data were recorded with a sampling period $Ts=0.005$ ($s$), generally resulting in a dataset with approximately \(12,000\) samples (60 $s$) at most.

The programming language used in this research was Matlab\textsuperscript{\tiny\textregistered}, both for developing the optimization algorithm and for the system identification procedure via NNs. The simulation system was developed using the Matlab/Simulink\textsuperscript{\tiny\textregistered} platform, which contains both sine and step trajectories, a saturation block to bound the control signal from $0\;\mu s$ to $\rho_{max}\;\mu s$ for each individual, the RISE controller block, and the identified NN block for each individual.

\subsubsection{More than one session}
\label{subsec:two_five_sessions}

\textcolor{black}{For individuals who participated in more sessions, with at least 48 hours of difference between two consecutive sessions, the one-minute stimulation test (identification step) was not considered, as it was performed when an individual participated for the first time.} The data from previous rehabilitation sessions were used to train an NN model in an offline scheme. Before a new session, all data from an individual were combined to a single dataset and used to better map the relationship between angular position and electrical stimulation. In this study, we only used the control data resulting from the control-stimulation sessions with fine-tuned IGA gains, as it would be in real life rather than empirical gains.

Thus, using each trained identified model, we performed an optimization procedure based on the IGA to find the best gain combination for both sine and step reference trajectories. As this optimization was performed in an offline scheme and before the next session, time and computational costs were not too strict as they were for the first session. Therefore, the initial parameters of IGA used for simulations were as follows: population size $N_p=10$, mutation rate \(M_r=0.3\), number of generations \(N_g=30\) (size of RIP) and \(k=1\) iteration. The algorithm ran only once, and several gain combinations from the set of solutions were simulated to check the system response and select the best gain combination for both trajectories. \textcolor{black}{Generally, the total time for both system identification and RISE gains optimization procedures took about 1 $h$ for each individual/session.}

For the experimental part, the electrodes were positioned at the motor-point identified in the first session, and similarly, a few open-loop tests, applying a step input during four seconds, were performed, to determine a bounded PW band related to $\theta_{min}=10\degree$ and $\theta_{max}=40\degree$. Afterward, a small-time interval for muscle rest was provided.

Therefore, knowing the fine-tuned gain parameters for each individual, we applied the controlling procedure for both references, and then employed an empirical gain combination for comparing results. 

\section{Results and Discussion}
\label{sec:resul}
In this section, we report the results obtained by applying our proposed methodology in real experiments. During this study, individuals H1-H4 participated in five sessions, H5 in three sessions, and H6-H7 in two sessions. Individuals with SCI participated in only one session due to displacement difficulties. For all individuals, the first session took more time and one additional stimulation than the subsequent ones. This was due to the one-minute stimulation test, and the training/optimization time during the session to find the best gain combination. Before the start of any control-stimulation test, five combinations of empirical gains (\(\alpha_1;\alpha_2;ks;\beta\)) were chosen as (\(1;2;30;5\)), (\(0.5;1;30;1.5\)), (\(0.8;1.2;20.5;2.5\)), (\(5;2;15;3\)), (\(4;7;25;8\)) for sessions one to five, respectively. As the system responses to any combination of gains were unknown, they were all chosen at random. Subsections~\ref{sub:res_experiments} and \ref{sub:res_ident} present our nonlinear control and system identification results and analysis, respectively. Lastly, we provide a general discussion in Subsection~\ref{sub:discussion}.

\subsection{Control-based NMES results}\label{sub:res_experiments}

\begin{figure}[!h]
	\centering
	\includegraphics[width=1\linewidth]{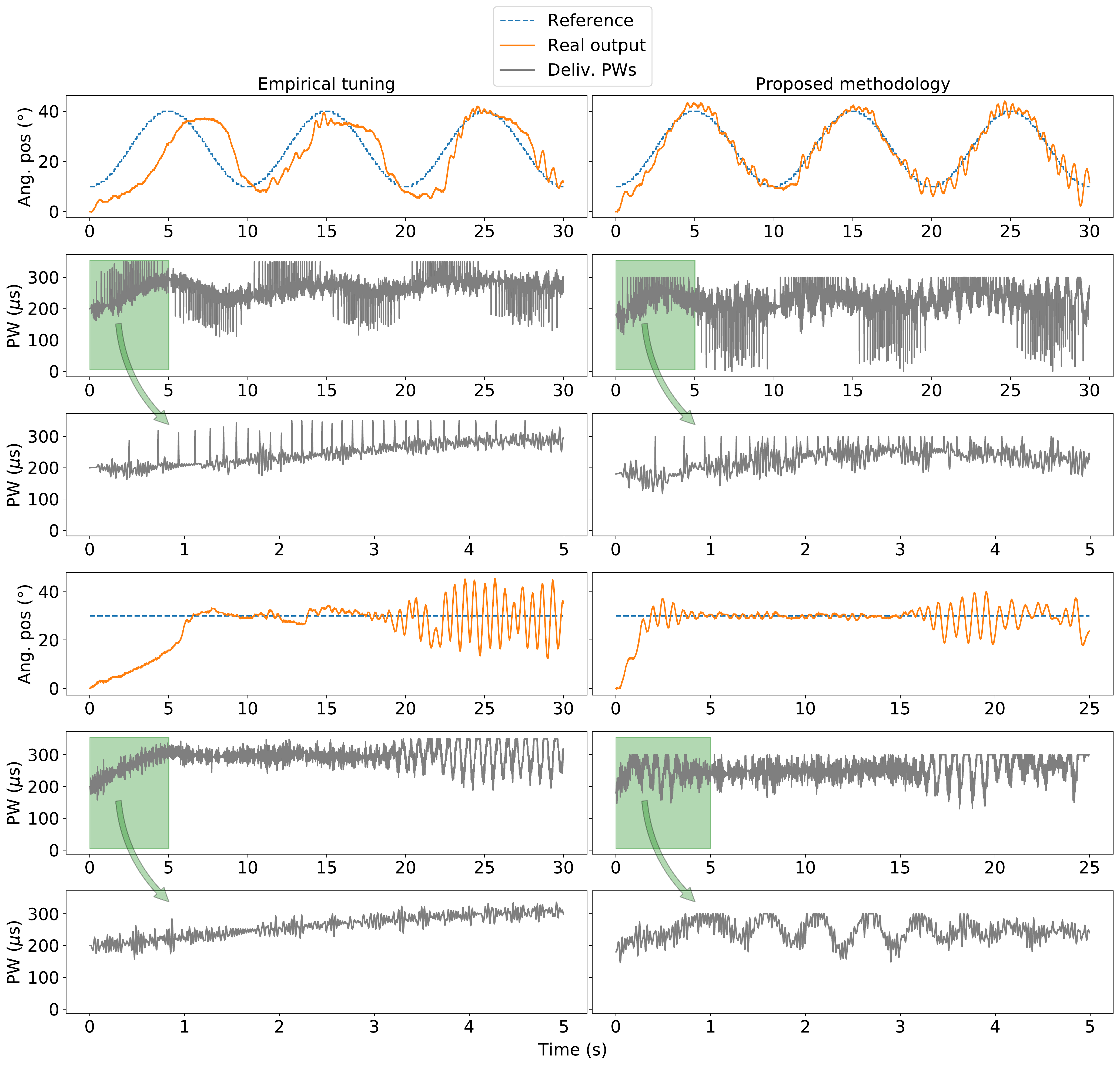}
	\caption{\textcolor{black}{Experimental results for individual P1 comparing empirical gains and the proposed methodology. The first and second rows illustrate the tracking results for the sine wave and the corresponding delivered PWs (with zoom during five seconds on the third row), respectively. Similarly, the fourth and fifth rows illustrate the tracking results for the step wave and the corresponding delivered PWs (with zoom during five seconds on the last row), respectively.}}
	\label{fig:p1_results}
\end{figure}

\begin{figure}[!h]
	\centering
	\includegraphics[width=1\linewidth]{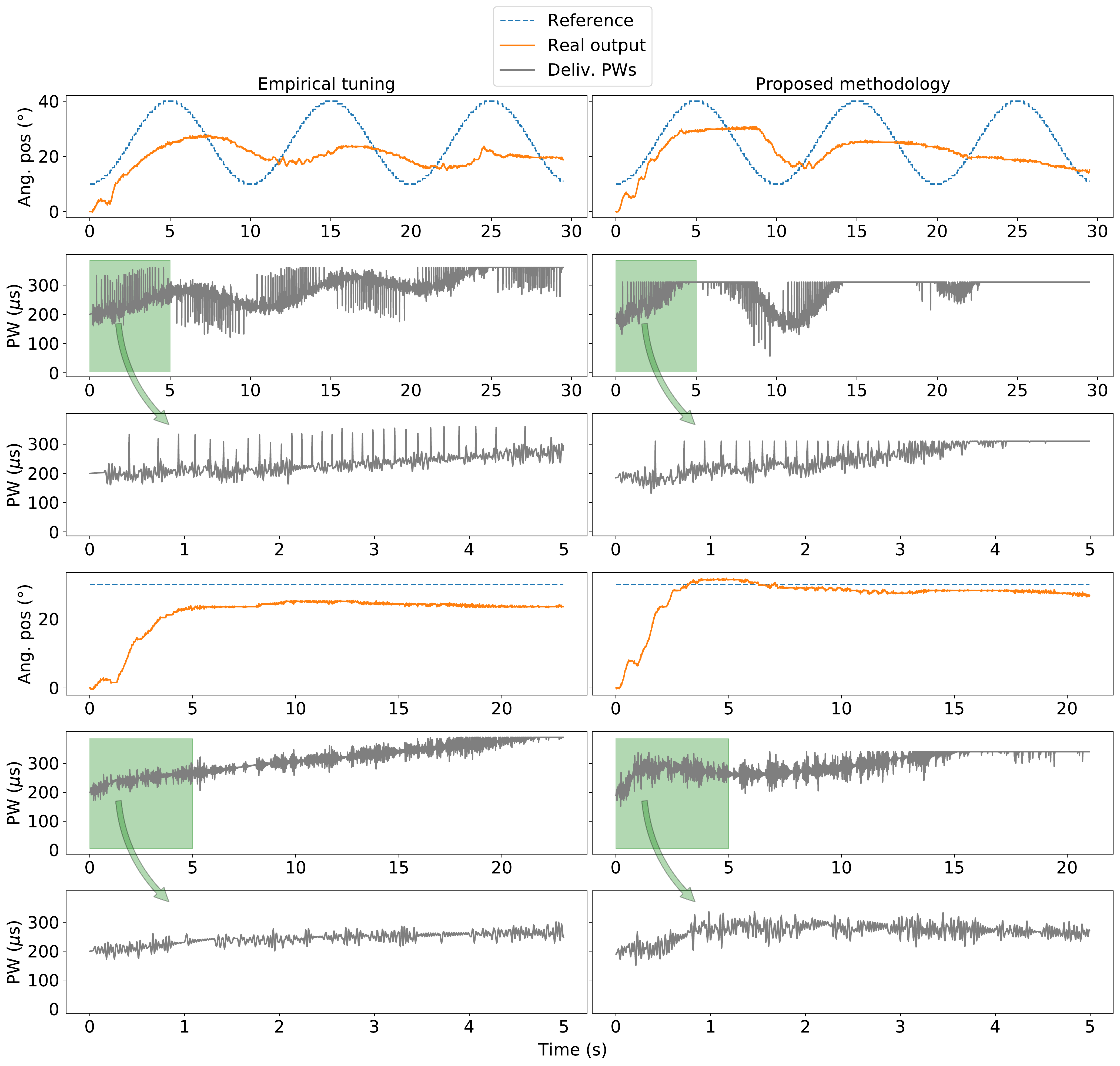}
	\caption{\textcolor{black}{Experimental results for individual P2 comparing empirical gains and the proposed methodology. The first and second rows illustrate the tracking results for the sine wave and the corresponding delivered PWs (with zoom during five seconds on the third row), respectively. Similarly, the fourth and fifth rows illustrate the tracking results for the step wave and the corresponding delivered PWs (with zoom during five seconds on the last row), respectively.}}
	\label{fig:p2_results}
\end{figure}

Figures~\ref{fig:p1_results} and~\ref{fig:p2_results} illustrate the tracking results on both trajectories and their delivered PWs (Deliv. PWs) for individuals P1 and P2, respectively. Additionally, \textcolor{black}{Table~\ref{tbl_rmse} presents control results for the sine wave}, comparing the proposed methodology with an empirical tuning for all individuals (\textit{Ind.}) in each session (\textit{Sess.}). The metrics in this table are the root mean square error (RMSE) between the desired and actual knee angles considering the whole period of control-stimulation; and the time of effective control (TEC), which represents how much time in seconds the lower limb was control-stimulated to track the reference angle. When the lower limb did not track the reference angle, the RMSE metric is represented by NC, meaning \enquote{not calculated}. More precisely, the TEC metric is the time between the initial control-stimulation until the leg stops tracking the reference angle ($\pm 5\degree$ error) for 5 $s$. In the worst-case, if the leg never tracks the reference angle, NC is assigned. 

\textcolor{black}{Similarly, Table~\ref{tbl_mean_gains} presents control results for the step wave, comparing the proposed methodology with an empirical tuning for all individuals in each session. The metrics in this table are the RMSE, TEC, and the averaged and standard deviation (std) values of the knee angular position around the operating point (AvStd. OP) in degrees. The \textit{AvStd. OP} metric will be regarded as an indicator to evaluate the oscillatory behavior during regulation around an operation point ($40\degree$ for healthy individuals and $30\degree$ for individuals with SCI). For individuals who participated in more than one session, Tables~\ref{tbl_rmse} and~\ref{tbl_mean_gains} present the averaged (Avg.) and the std values for both RMSE and TEC metrics, which are calculated considering all sessions of each individual. The symbol (*) indicates there is no std value, as there are more ``NC" than real values.} Lastly, similar to Figs.~\ref{fig:p1_results} and~\ref{fig:p2_results}, \ref{app:sup_results} provides supplementary illustrations for the tracking results of individuals H1 (session \textit{v}), H2 (session \textit{ii}), and H4 (session \textit{v}), respectively, \textcolor{black}{as well as the fine-tuned gains (\(\alpha_1\); \(\alpha_2\); \(ks\); \(\beta\)) used for each RISE-based control-stimulation session, in Table~\ref{tbl_gains}.}

\begin{table}[!htb]
    \centering
    \scriptsize
    \caption{\textcolor{black}{Performance results for the sine wave on control experiments using the proposed methodology and empirical tuning for all individuals in their respective sessions.}}
    \label{tbl_rmse}
    \scalebox{0.9}{
    \begin{tabular}{c c c c c c}
    \toprule
    \multirow{2}{*}{\textbf{Ind.}} &\multirow{2}{*}{\textbf{Sess.}} &\multicolumn{2}{c}{\textbf{Empirical}} &\multicolumn{2}{c}{\textbf{Proposed methodology}} \\\cmidrule{3-6}
    & &\textbf{RMSE} &\textbf{TEC} &\textbf{RMSE} &\textbf{TEC}\\
    \midrule
    P1 &i &9.147\degree &30 $s$ &2.984\degree &30 $s$ \\\hline
    P2 &i &11.296\degree &30 $s$ &10.730\degree &30 $s$ \\\hline
    \multirow{5}{*}{H1} &i &7.494\degree &60 $s$ &5.830\degree &60 $s$ \\
    &ii &8.752\degree &60 $s$ &5.933\degree &60 $s$  \\
    &iii &14.092\degree &60 $s$ &7.337\degree &60 $s$ \\
    &iv &6.377\degree &60 $s$ &3.629\degree &60 $s$  \\
    &v &6.383\degree &60 $s$ &3.562\degree &60 $s$  \\
    \multicolumn{2}{c}{\textbf{Avg.(std)}} &8.62(2.87)\degree   &60(0) $s$    &5.26(1.46)\degree   &60(0) $s$ \\\midrule
    \multirow{5}{*}{H2} &i &5.212\degree &60 $s$ &5.055\degree &45 $s$ \\
    &ii &8.317\degree &60 $s$ &3.885\degree &60 $s$  \\
    &iii &11.741\degree &35 $s$ &3.633\degree &60 $s$  \\
    &iv &4.887\degree &40 $s$ &3.562\degree &40 $s$   \\
    &v &10.713\degree &33 $s$ &4.858\degree &23 $s$  \\
    \multicolumn{2}{c}{\textbf{Avg.(std)}}  &8.17(2.78)\degree &46(12) $s$ &4.20(0.63)\degree  &46(14) $s$ \\\midrule
    \multirow{5}{*}{H3} &i &NC &NC &6.019\degree &30 $s$ \\
    &ii &9.221\degree &50 $s$ &7.615\degree &50 $s$ \\
    &iii &NC &NC &4.616\degree &33 $s$  \\
    &iv &3.775\degree &33 $s$ &6.688\degree &60 $s$\\
    &v &19.096\degree &60 $s$ &6.516\degree &60 $s$  \\
    \multicolumn{2}{c}{\textbf{Avg.(std)}} &10.70(6.3)\degree  &48(11) $s$  &6.29(0.98)\degree    &47(13)$s$  \\\midrule
    \multirow{5}{*}{H4} &i &NC &NC &9.382\degree &60 $s$\\
    &ii &12.794\degree &60 $s$ &4.823\degree &60 $s$ \\
    &iii  &8.246\degree &60 $s$ &4.640\degree &30 $s$ \\
    &iv &3.534\degree &31 $s$ &4.561\degree &60 $s$\\
    &v &16.483\degree &60 $s$ &3.717\degree &42 $s$ \\
    \multicolumn{2}{c}{\textbf{Avg.(std)}}  &10.3(4.86)\degree &53(13) $s$ & 5.42(2.01)\degree  &50(12) $s$  \\\midrule
    \multirow{3}{*}{H5} &i &NC &NC &6.006\degree &20 $s$ \\
    &ii &8.070\degree &50 $s$ &3.017\degree &21 $s$   \\
    &iii &NC &NC &3.872\degree &52 $s$ \\
    \multicolumn{2}{c}{\textbf{Avg.(std)}}  &8.070(*)\degree  &50(*) $s$ & 4.30(1.26)\degree    &31(15)$s$  \\\midrule
    \multirow{2}{*}{H6} &i &NC &NC &10.128\degree &60 $s$ \\
    &ii &9.105\degree &60 $s$ &6.553\degree &60 $s$ \\
    \multicolumn{2}{c}{\textbf{Avg.(std)}} &9.105(*)\degree  &60(*) $s$  &8.34(1.79)\degree  &60(0)$s$ \\\midrule
    \multirow{2}{*}{H7} &i &NC &NC &8.500\degree &60 $s$ \\
    &ii &NC &NC &6.630\degree &50 $s$ \\
    \multicolumn{2}{c}{\textbf{Avg.(std)}}  &NC &NC &7.56(0.94)\degree  &55(5) $s$ \\
    \bottomrule
    \end{tabular}}
\end{table}

\begin{table}[!h]
    \centering
    \scriptsize
    \caption{\textcolor{black}{Performance results for the step wave on control experiments using the proposed methodology and empirical tuning for all individuals in their respective sessions.}}
    \label{tbl_mean_gains}
    \scalebox{0.9}{
    \begin{tabular}{c c c c c c c c}
    \toprule
    \multirow{2}{*}{\textbf{Ind.}} &\multirow{2}{*}{\textbf{Sess.}} &\multicolumn{3}{c}{\textbf{Empirical}} &\multicolumn{3}{c}{\textbf{Proposed methodology}} \\\cmidrule{3-8}
    & &\textbf{RMSE} &\textbf{TEC} &\textbf{AvStd. OP} &\textbf{RMSE} &\textbf{TEC} &\textbf{AvStd. OP} \\
    \midrule
    P1 &i &10.995\degree &30 $s$ &29.44(6.03)\degree  &5.978\degree &25 $s$ &29.88(3.25)\degree  \\\hline
    P2 &i &10.106\degree &23 $s$ &24.15(0.59)\degree  &6.613\degree &21 $s$ &28.51(0.99)\degree \\\hline
    \multirow{5}{*}{H1} &i  &5.920\degree &60 $s$ & 39.35(2.50)\degree  &6.167\degree &60 $s$  & 39.54(1.63)\degree \\
    &ii  &12.291\degree &60 $s$ & 36.19(6.92)\degree & 8.201\degree &60 $s$  & 39.54(4.61)\degree\\
    &iii  &7.266\degree &60 $s$ & 37.61(3.76)\degree & 4.164\degree &60 $s$  & 39.98(1.48)\degree  \\
    &iv  &6.741\degree &60 $s$ & 38.70(4.59)\degree &4.404\degree &60 $s$   & 39.89(1.64)\degree \\
    &v  &6.887\degree &60 $s$ & 39.94(5.88)\degree  &4.425\degree &60 $s$    & 40.01(1.34)\degree \\
    \multicolumn{2}{c}{\textbf{Avg.(std)}} &7.82(2.28)\degree   &60(0) $s$  & - &5.47(1.54)\degree   &60(0) $s$  & - \\\midrule
    \multirow{5}{*}{H2} &i &9.764\degree &35 $s$ & 38.45(2.00)\degree  &6.212\degree &37 $s$     & 39.80(2.50)\degree   \\
    &ii &NC &NC &NC  &7.856\degree &25 $s$    & 38.05(1.53)\degree \\
    &iii &11.822\degree &57 $s$ & 33.50(4.14)\degree  &5.457\degree &37 $s$    & 39.88(3.36)\degree   \\
    &iv  &6.424\degree &34 $s$ & 38.94(1.92)\degree  &4.890\degree &45 $s$    & 39.42(1.33)\degree \\
    &v  &6.226\degree &35 $s$ & 39.83(4.11)\degree &7.233\degree &38 $s$  & 40.19(3.34)\degree \\
    \multicolumn{2}{c}{\textbf{Avg.(std)}}  &8.56(2.35)\degree &46(11) $s$ &  -  &6.33(1.09)\degree  &37(6) $s$ & -    \\\midrule
    \multirow{5}{*}{H3} &i  &15.359\degree &48 $s$ & 32.47(7.59)\degree  &8.176\degree &32 $s$  & 39.54(1.34)\degree \\
    &ii  &8.230\degree &45 $s$ & 38.53(2.19)\degree &5.598\degree &28 $s$  & 39.71(0.63)\degree \\
    &iii  &14.233\degree &38 $s$ & 33.06(6.26)\degree   &6.258\degree &30 $s$  & 39.66(0.95)\degree \\
    &iv   &5.472\degree &40 $s$ & 39.54(0.71)\degree &6.357\degree &55 $s$  & 39.64(5.21)\degree \\
    &v  &7.102\degree &60 $s$ & 39.88(6.52)\degree  &4.491\degree &60 $s$   & 39.84(2.23)\degree\\
    \multicolumn{2}{c}{\textbf{Avg.(std)}} &10.08(3.97)\degree  &46(8) $s$ & - &6.18(1.2)\degree    &41(14) $s$ & -\\\midrule
    \multirow{5}{*}{H4} &i  &13.914\degree &60 $s$ & 39.08(9.81)\degree &5.943\degree &60 $s$   & 40.02(2.97)\degree \\
    &ii  &8.354\degree &60 $s$ & 40.49(2.82)\degree  &4.694\degree &60 $s$ & 40.00(0.87)\degree\\
    &iii   &8.830\degree &60 $s$ & 42.26(1.86)\degree   &7.286\degree &60 $s$  & 39.92(2.35)\degree\\
    &iv   &4.551\degree &60 $s$ & 39.89(1.47)\degree    &6.777\degree &60 $s$  & 39.82(5.33)\degree \\
    &v &7.871\degree &60 $s$ & 39.92(7.29)\degree  &4.895\degree &60 $s$   & 39.88(2.52)\degree \\
    \multicolumn{2}{c}{\textbf{Avg.(std)}} & 8.70(3.01)\degree &60(0) $s$ & - & 5.92(1.02)\degree  &60(0) $s$ &  - \\\midrule
    \multirow{3}{*}{H5} &i &NC &NC &NC   &5.719\degree &60 $s$    &  39.83(3.74)\degree \\
    &ii  &8.076\degree &52 $s$ & 39.13(2.43)\degree   &5.481\degree &50 $s$  & 39.58(1.11)\degree  \\
    &iii &13.032\degree &45 $s$ & 33.66(5.39)\degree   &6.351\degree &50 $s$    & 39.42(1.99)\degree  \\
    \multicolumn{2}{c}{\textbf{Avg.(std)}}   &10.55(2.48)\degree  &47(3) $s$ & - & 5.85(0.37)\degree    &57(5)$s$    & -  \\\midrule
    \multirow{2}{*}{H6} &i  &12.789\degree &60 $s$ & 31.64(5.37)\degree   &6.578\degree &60 $s$   & 40.01(2.80)\degree \\
    &ii &7.506\degree &60 $s$ & 39.6(2.89)\degree    &4.040\degree &60 $s$  & 39.62(1.49)\degree \\
    \multicolumn{2}{c}{\textbf{Avg.(std)}} &10.15(2.64)\degree  &60(0) $s$ & - &5.31(1.27)\degree     &60(0) $s$  & - \\\midrule
    \multirow{2}{*}{H7} &i  &9.554\degree &40 $s$ & 38.82(4.25)\degree   &7.044\degree &21 $s$  & 39.63(2.49)\degree \\
    &ii &13.135\degree &60 $s$ & 36.38(10.51)\degree    &5.212\degree &60 $s$    & 40.04(1.68)\degree \\
    \multicolumn{2}{c}{\textbf{Avg.(std)}}  &11.34(1.79)\degree &50(10) $s$ & - &6.13(0.92)\degree &40(19) $s$ & - \\
    \bottomrule
    \end{tabular}}
\end{table} 

As shown in Tables~\ref{tbl_rmse} and~\ref{tbl_mean_gains} and Figs.~\ref{fig:p1_results} and~\ref{fig:p2_results}, the proposed methodology could be effectively applied to clinical procedures for treating people with SCI via NMES/FES. In general, tremors (mainly for P1) and fatigue were detected for both individuals with SCI at the end of each trajectory (sine and wave). This was because neither of them had been admitted to a rehabilitation research program involving daily electrical stimulation exercise of their lower limbs. In all experiments, P1 had no perception of the stimulation, while P2 experienced small discomfort due to the electrical stimulation intensity. Results from P1 validate and substantiate the first hypothesis presenting very good tracking results using the proposed methodology. When empirical gains were used, the lower limb tracked the sine wave with a lag and presented a slow response to the step trajectory. Moreover, the RMSE of \(2.9842\degree\) for the IGA sine wave from P1 was the best result during all experiments in this research, which is a third of the RMSE obtained using empirical gains \(9.1471\degree\). However, in the final seconds (about 28 $s$), the lower limb would start to have more tremors due to the fatigue factor; this is also noticed after about 15 $s$ to the step wave for both our proposed method and empirical tuning.

Furthermore, the tracking result for the sine wave of individual P2 was not as satisfactory such as for P1. However, as seen in the Deliv. PWs curve (Fig.~\ref{fig:p2_results}), this poor sine wave tracking could be due to an underestimation for the upper bound to the control signal value (as this individual experienced discomfort under NMES); selecting a higher value may have resulted in good tracking. This inference is substantiated by the good results achieved in the step wave after a 3 min interval for muscle rest and by having consent to increase the upper bound value to the PW. A good regulation around the operation point was achieved for approximately 21 $s$, with $50\%$ less RMSE than that obtained using empirical tuning. More specifically, using empirical tuning led to poor performances for both sine and step trajectories, as the leg did not track the sine wave, and the regulation around the operation point featured a stationary error.


For healthy individuals, as seen in Tables~\ref{tbl_rmse} and~\ref{tbl_mean_gains} and in the figures of \ref{app:sup_results}, using empirical gains led to several poor performances. In many tests, the control-stimulated lower limb did not track the reference angle (\enquote{NC}) or presented high oscillatory comportment. \textcolor{black}{This problem is, for example, demonstrated in Figs.~\ref{fig:h1_results} and~\ref{fig:h4_results} and in Table~\ref{tbl_mean_gains} regarding the \textit{AvStd. OP} metric, as using empirical gains resulted in average values (knee angular position) below the operation point with high std values.} When the proposed methodology was used, for all individuals, satisfactory and suitable tracking results were acquired for both the tracking of sine wave via isotonic contraction and the regulation around an operation point (step wave) as isometric contraction. On average, for each individual, our proposed methodology presented much lower RMSE while still achieving high TEC (Tables~\ref{tbl_rmse} and~\ref{tbl_mean_gains}). 

Finally, for most healthy individuals, when our proposed solution was used and RISE controller was not tuned with empirical gains, the lower limb robustly tried to track the reference angle for 60 $s$. This could not be possible if we had performed pretrial tests, which could generate muscle fatigue due to prior stimulations. In contrast, RISE controller presented by~\cite{Stegath2007} and \cite{Stegath2008} demonstrated tracking control for 8 $s$ for a step trajectory and 20 $s$ for a sine wave;~\cite{Sharma2009} and~\cite{Sharma2012} presented tracking control for 30 $s$ for a step- and a sine-type signal; \cite{Kushima2015} presented tracking control for 30 $s$ for a sine wave, and~\cite{Downey2015_b} presented tracking control for 45 $s$ (for conventional stimulation) for a sine trajectory. \textcolor{black}{On the other hand, in some of our experiments, significant ``chattering" was noticed in the control input. Yet, except for individual P2, none of the other voluntary participants reported discomfort due to NMES while presenting satisfactory tracking of the lower limb with high TEC. Further improvements to RISE controller tuning (i.e., IGA) may help to smooth this ``chattering" problem, which is undesirable and may lead to poor controller performance (\cite{Lynch2012})}.

\subsection{Nonlinear system identification results}\label{sub:res_ident}
Table~\ref{tbl_identification} presents the following metrics for all individuals (\textit{Ind.}) in each session (\textit{Sess.}): (i) the Pearson correlation coefficient (\textit{Corr.}) between the input (PW) and output (angular position) data using past control data as sessions progress; (ii) the Coefficient of determination ($R^2$); and (iii) the mean squared error (\textit{MSE}). These metrics are explained in the following: First, the \textit{Corr.}  between the input and output data indicates the correlation between both data, which clarifies how ``difficult" it is to identify the system dynamics. More specifically, \textit{Corr.} measures the linear correlation between two variables \textit{x} and \textit{y}. The \textit{Corr.} value ranges from -1 to 1. The higher the value, the stronger the correlation. A negative value indicates an inverse correlation, while a positive value indicates a regular correlation. Second, the coefficient of determination ($R^2$) is the proportion of the variance in the dependent variable that is predictable from the independent variable. The larger $R^2$ is, the more the variability is indicated by the linear regression model\footnote{https://fr.mathworks.com/help/stats/coefficient-of-determination-r-squared.html}. Third, the MSE is the average squared error between the NN outputs and the real ones.

\begin{figure}[!htb]
	\centering
	\includegraphics[width=1\linewidth]{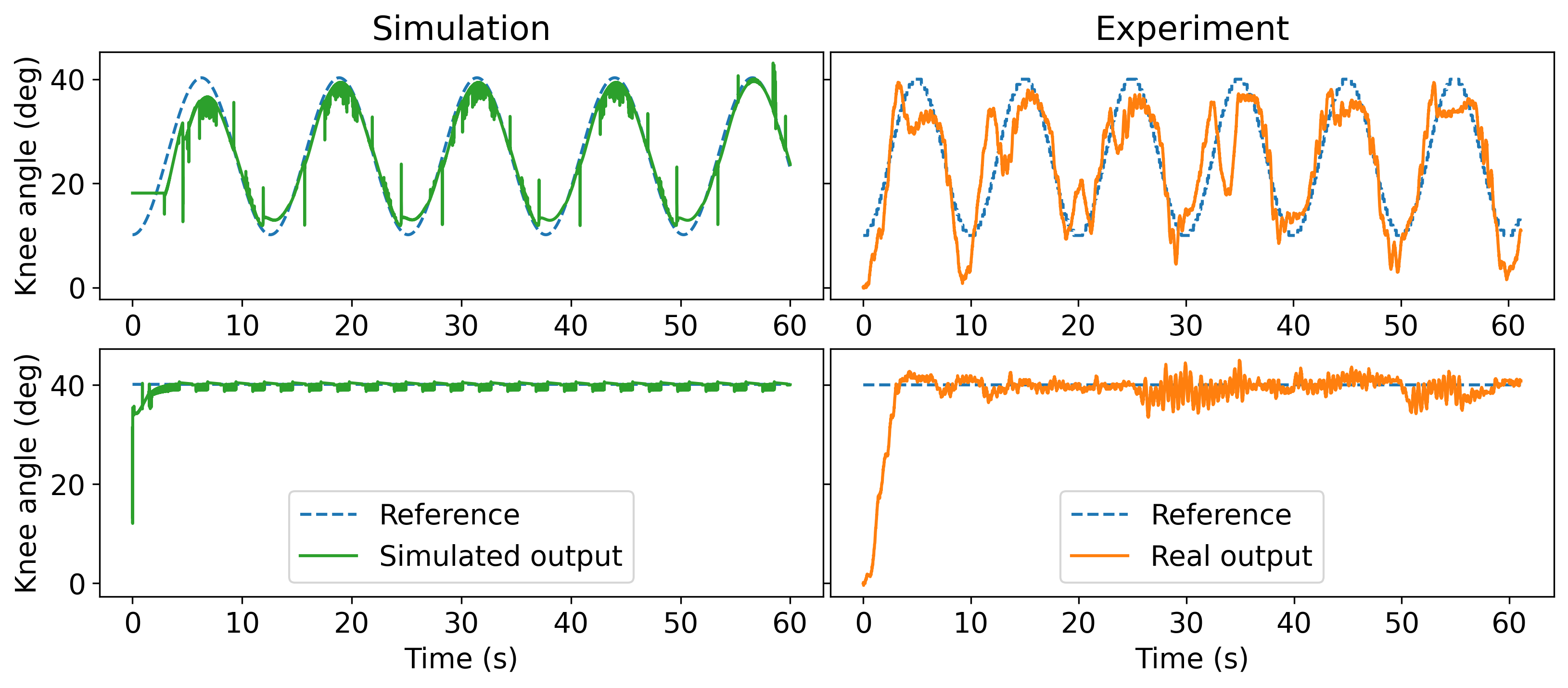}
	\caption{Comparison of simulation and real experiments for individual H1 using past rehabilitation data to identify the nonlinear model.}
	\label{fig:H1_simVSreal}
\end{figure}

\begin{figure}[!htb]
	\centering
	\includegraphics[width=1\linewidth]{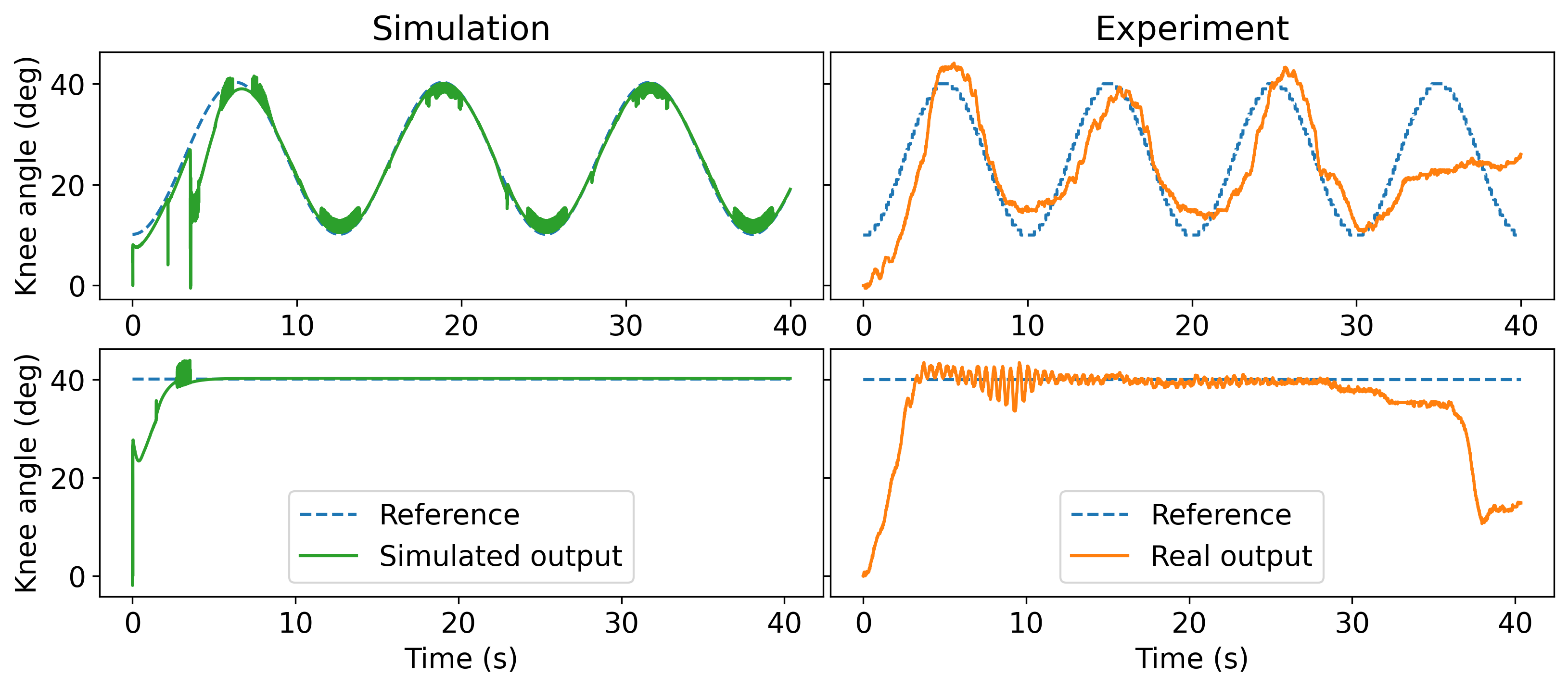}
	\caption{Comparison of simulation and real experiments for individual H3 using past rehabilitation data to identify the nonlinear model.}
	\label{fig:H3_simVSreal}
\end{figure}

Additionally, Figs.~\ref{fig:H1_simVSreal} and~\ref{fig:H3_simVSreal} compare the results from simulation and real experiments using either empirical or fine-tuned IGA gains. These figures were selected for illustration purposes only, as the objective is to highlight the benefits of using past data for the nonlinear system identification step. In \ref{app:sup_results}, we provide more illustrations comparing simulation versus the real experiment results.

\begin{table}[!htb]
    \centering
    \scriptsize
    \caption{Identification results for all individuals in their respective sessions.}
    \label{tbl_identification}
    \begin{tabular}{c c c c c| c c c c c}
    \hline
    \textbf{Ind.} &\textbf{Sess.} &\textbf{Corr.} &\textbf{$R^2$} &\textbf{MSE} &\textbf{Ind.} &\textbf{Sess.} &\textbf{Corr.} &\textbf{$R^2$} &\textbf{MSE}\\
    \hline
    P1 &i &0.4153 &0.836 &0.001     &P2 &i &0.1035 &0.796 &0.003  \\\hline
    \multirow{5}{*}{H1} &i &0.5908 &0.726 &0.002    &\multirow{5}{*}{H2}&i &0.7789 &0.869 &0.006 \\
    &ii &0.1738  &0.157 &0.038 &                                          &ii    &0.2594  &0.416 &0.039 \\
    &iii &0.0469  &0.101 &0.042 &                                             &iii    &0.2640  &0.308 &0.039 \\
    &iv &-0.1109  &0.159 &0.041 &                                          &iv    &0.2606  &0.282 &0.037 \\
    &v &-0.0916  &0.113 &0.040 &                                           &v    &0.2769  &0.292 &0.038 \\\hline
    \multirow{5}{*}{H3} &i &0.8333  &0.820 &0.003 &\multirow{5}{*}{H4}&i &0.7339 &0.974 &0.001 \\
    &ii &0.4325  &0.498 &0.023 &                              &ii   &0.0476   &0.377 &0.054 \\
    &iii &0.3083  &0.506 &0.023  &                            &iii  &-0.1050    &0.323 &0.054 \\
    &iv &0.3523  &0.510 &0.024 &                              &iv  &-0.0502    &0.294 &0.052 \\
    &v &0.2650  &0.492 &0.026 &                           &v  & -0.1017  &0.281 &0.049 \\\hline
    \multirow{3}{*}{H5} &i &0.6738  &0.881 &0.001 &\multirow{3}{*}{H6} &i &0.7182  &0.815 &0.001 \\
    &ii &0.3774  &0.682 &0.030                                &&ii  &0.3078  &0.476 &0.028\\
    &iii &0.3458  &0.599 &0.035 & & - & - & - &-  \\\hline
    \multirow{2}{*}{H7} &i &0.5201  &0.767 &0.004  &\multirow{2}{*}{-} & - & - & - &- \\
    &ii &0.5520  &0.476	   &0.017 & & - & - & - &-  \\
    \hline
    \end{tabular}
\end{table} 

As presented in Table~\ref{tbl_identification}, there are considerable decrements in the \textit{Corr.} between the input and output data as sessions progress (given the addition of control data from every new session). Moreover, while the data from healthy individuals in the first session are highly correlated ($0.5201 \leq Corr. \leq 0.8333$), the ones from individuals with SCI are poorly correlated, as their muscles do not respond to NMES/FES as well as the muscles of the healthy individuals. Moreover, due to less correlation between data, the generalization and learning procedure of an NN is harder, which resulted in increments in the error metrics $R^2$ and \textit{MSE}. However, as shown in the Figs.~\ref{fig:H1_simVSreal} and~\ref{fig:H3_simVSreal}, these models better describe what happens in real experiments, where non-ideal conditions such as fatigue, tremors, and spasms, are explained by data.

As shown in Figs.~\ref{fig:H1_simVSreal} and~\ref{fig:H3_simVSreal}, the identified models simulated a sine trajectory with some tremors in the upper and lower peak values and some tremors to the step wave around the operation point. These behaviors were also noticed in real experiments. \ref{app:sup_results} presents simulations to the step wave with a quick response and with oscillatory behavior for the whole period, which was also verified in real experiments. Although not flawless, such models can provide more insights into the real system response. However, as approximate models of healthy individuals, neither of them would be suitable for an exact description of the real system; for instance, voluntary movements, fear, or other issues related to the individual thoughts (e.g., social or personal life), can affect results, and these aspects could not be predicted by the model.
 
\subsection{Discussion}\label{sub:discussion}

\textcolor{black}{Because this research was conducted with volunteering participants, we depended on their availability. For instance, not all healthy individuals participated in the pre-established five sessions. One volunteer showed availability to participate in only two sessions (H7), two volunteers showed availability to participate in only three sessions (H5 and H6), while the others (H1-H4) participated in all five sessions. This way, rather than excluding the non-uniform collected data, we preferred to present our results for all volunteers according to the number of sessions they participated in. Note that this procedure is common in studies in this area, because each session depends on consent, as established by the ethics committee.}

\begin{figure}[!htb]
	\centering
	\includegraphics[width=1\linewidth]{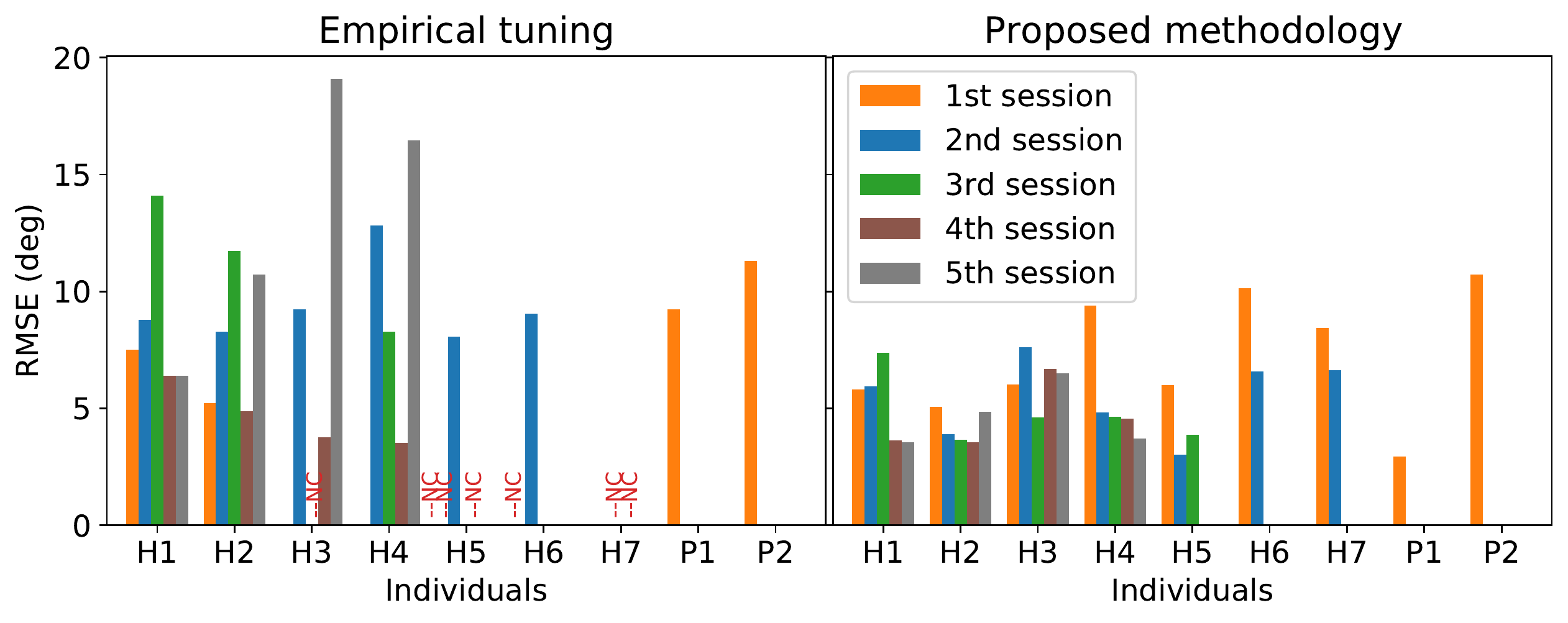}
	\caption{RMSE analysis for the sine trajectory based on empirical tuning or the proposed methodology.}
	\label{fig:rmse_sine}
\end{figure}

\begin{figure}[!htb]
	\centering
	\includegraphics[width=1\linewidth]{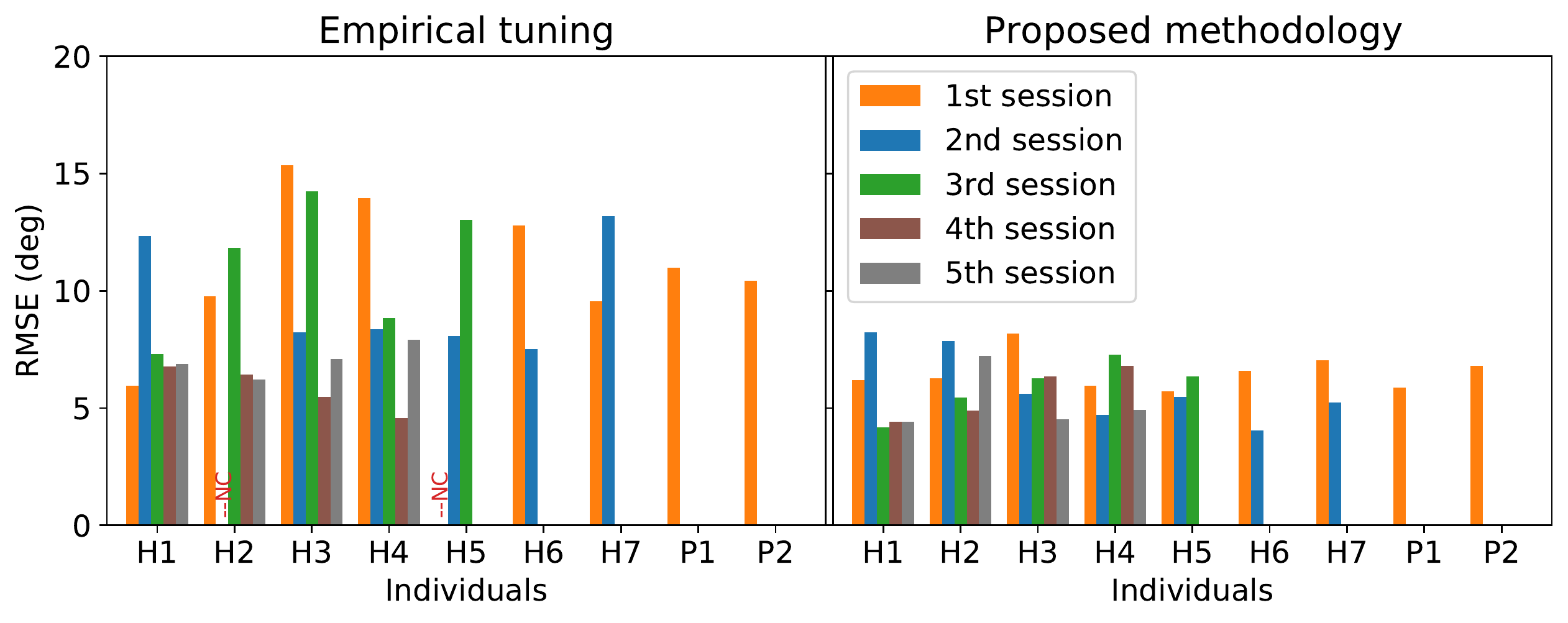}
	\caption{RMSE analysis for the step trajectory based on empirical tuning or the proposed methodology.}
	\label{fig:rmse_step}
\end{figure}

Figures~\ref{fig:rmse_sine} and~\ref{fig:rmse_step} summarize the results of Tables~\ref{tbl_rmse} and~\ref{tbl_mean_gains} by illustrating in bar plots the RMSE metric for both the empirical tuning and our proposed methodology, considering each trajectory (sine and step) in all sessions and all individuals. In these figures, \enquote{NC} indicates when the leg did not track the reference angle. Omitted bars indicate the individual did not participate in the corresponding session.

As demonstrated in Figs.~\ref{fig:rmse_sine} and~\ref{fig:rmse_step}, the proposed methodology consistently and considerably outperforms the empirical tuning approach, which supports and validates the first hypothesis made in this paper. Additionally, in the first sessions of healthy individuals, the RMSEs were generally higher, which could be due to fear or discomfort to the electrical stimulation or voluntary movements. However, as sessions progressed, the tracking results improved for some individuals, which among many factors, can be explained by the use of a more representative model with past rehabilitation data. This resulted in a better tuning of the RISE controller improving the tracking results in practice.

More precisely, considering the first hypothesis made in this paper, setting empirical gains to RISE controller generally led to an underperformance compared with the use of ML-based algorithms to find the best combination for each individual. That is, to provide efficient treatment for individuals with SCI via NMES/FES, a fine-tuning method such as the presented methodology can prevent SCI patients from experiencing premature fatigue and other problems during rehabilitation.

Moreover, for the second hypothesis, the use of the past rehabilitation data for the nonlinear system identification task also presented promising results. Even though there is less correlation between the input and output data, which increases the error on the identification process, the identified models ``gained" implicit non-ideal conditions such as tremors and spasms (\textit{cf.}, Figs.~\ref{fig:H1_simVSreal} and~\ref{fig:H3_simVSreal} and figures in~\ref{app:sup_results}). Therefore, using data from past rehabilitation sessions of each individual and strong tools, such as NNs, the mapping over the delivered electrical stimulation and the angular position can be efficiently addressed with more realistic models. Regarding future work, we recommend and intend to explore a deeper comparison between a case considering past rehabilitation data and a case considering each session of an individual as the first one (applying and using only the one-minute stimulation test for system identification).

Finally, in the NMES-based knee simulation system, using data saved from each patient allows testing improvements to RISE control law and testing more control techniques before actual implementation, saving time and resources. Furthermore, the proposed methodology for knee joint control would allow people with no experience with technical information on neural networks, genetic algorithms, or even the control law RISE to easily use a closed-loop NMES/FES system for SCI individuals' rehabilitation.

\section{Conclusion}
\label{sec:conc}

Aiming to improve human lower limb tracking control of individuals with SCI via NMES/FES, this paper introduces a novel ML-based methodology. It consists of data-driven models that use past rehabilitation data, the RISE control method (or fundamentally similar control laws) to guarantee the semi-global asymptotic stability, and an improved genetic algorithm to efficiently tune the controller. Experiments were performed with seven healthy and two paraplegic individuals, which validated the proposed methodology. 

Additionally, RISE control method designed for lower limb control in the literature did not validate this RISE controller for paraplegic subjects. Therefore, for the first time and using the proposed methodology, we validated this controller with two SCI subjects with promising tracking results. This, however, would not be possible using a \enquote{trial and error} method by fatiguing the muscle before acquiring good tuning. Moreover, in the experiments performed in this research, for many healthy individuals, the lower limb robustly tried to track the reference angle for more than 45 $s$, which is the maximum time presented in the literature for RISE controller, reaching 60 $s$ many times.

We recommend and intend to explore the following areas for future work: a deeper validation with SCI patients under more sessions; a comparison of the effectiveness of using past rehabilitation data with different setups, e.g., the SCI patient is identified each session; the implementation of deeper and dynamic NNs studied in~\cite{Arcolezi2020} to improve identified models in our proposed methodology; \textcolor{black}{to improve the RISE controller tuning approach (i.e., IGA algorithm)}; the consideration of different control laws and improvements to RISE control method.

\section*{Acknowledgment}

\noindent This work was supported by the Coordenação de Aperfeiçoamento de Pessoal de Nível Superior - Brasil (CAPES) - Finance Code 001, by the Brazilian National Council for Scientific and Technological Development (CNPq) under research fellowships 309.872/2018-9 and 312.170/2018-1, and by the Region of Bourgogne Franche-Comté CADRAN Project. The authors would also like to thank each volunteer who participated in this research, especially those with spinal cord injury. We also thank anonymous reviewers for several comments that led to improvements in the paper as well as Enago for the English language review.


\bibliography{ms}

\begin{thebibliography}{53}
\expandafter\ifx\csname natexlab\endcsname\relax\def\natexlab#1{#1}\fi
\providecommand{\url}[1]{\texttt{#1}}
\providecommand{\href}[2]{#2}
\providecommand{\path}[1]{#1}
\providecommand{\DOIprefix}{doi:}
\providecommand{\ArXivprefix}{arXiv:}
\providecommand{\URLprefix}{URL: }
\providecommand{\Pubmedprefix}{pmid:}
\providecommand{\doi}[1]{\href{http://dx.doi.org/#1}{\path{#1}}}
\providecommand{\Pubmed}[1]{\href{pmid:#1}{\path{#1}}}
\providecommand{\bibinfo}[2]{#2}
\ifx\xfnm\relax \def\xfnm[#1]{\unskip,\space#1}\fi
\bibitem[{Arcolezi et~al.(2020)Arcolezi, Nunes, Cerna, de~Araujo, Sanches,
  Teixeira \& de~Carvalho}]{Arcolezi2020}
\bibinfo{author}{Arcolezi, H.~H.}, \bibinfo{author}{Nunes, W. R. B.~M.},
  \bibinfo{author}{Cerna, S.}, \bibinfo{author}{de~Araujo, R.~A.},
  \bibinfo{author}{Sanches, M. A.~A.}, \bibinfo{author}{Teixeira, M. C.~M.}, \&
  \bibinfo{author}{de~Carvalho, A.~A.} (\bibinfo{year}{2020}).
\newblock \bibinfo{title}{Identifying the knee joint angular position under
  neuromuscular electrical stimulation via long short-term memory neural
  networks}.
\newblock {\it \bibinfo{journal}{Research on Biomedical Engineering}\/}, .
  \DOIprefix\doi{10.1007/s42600-020-00089-1}.
\bibitem[{Arcolezi et~al.(2019)Arcolezi, Nunes, Nahuis, Sanches, Teixeira \&
  de~Carvalho}]{arcolezi2019}
\bibinfo{author}{Arcolezi, H.~H.}, \bibinfo{author}{Nunes, W. R. B.~M.},
  \bibinfo{author}{Nahuis, S. L.~C.}, \bibinfo{author}{Sanches, M. A.~A.},
  \bibinfo{author}{Teixeira, M. C.~M.}, \& \bibinfo{author}{de~Carvalho, A.~A.}
  (\bibinfo{year}{2019}).
\newblock \bibinfo{title}{A {RISE}-based controller fine-tuned by an improved
  genetic algorithm for human lower limb rehabilitation via neuromuscular
  electrical stimulation}.
\newblock In {\it \bibinfo{booktitle}{2019 6th International Conference on
  Control, Decision and Information Technologies ({CoDIT})}\/}.
\newblock \bibinfo{publisher}{{IEEE}}.
\newblock \DOIprefix\doi{10.1109/codit.2019.8820357}.
\bibitem[{Bergstra \& Bengio(2012)}]{Bergstra12randomsearch}
\bibinfo{author}{Bergstra, J.}, \& \bibinfo{author}{Bengio, Y.}
  (\bibinfo{year}{2012}).
\newblock \bibinfo{title}{Random search for hyper-parameter optimization}.
\newblock {\it \bibinfo{journal}{JMLR}\/},  (p. \bibinfo{pages}{305}).
\bibitem[{Bickenbach(2013)}]{who_sci}
\bibinfo{author}{Bickenbach, J.} (\bibinfo{year}{2013}).
\newblock {\it \bibinfo{title}{International perspectives on spinal cord
  injury}\/}.
\newblock \bibinfo{address}{Geneva, Switzerland}: \bibinfo{publisher}{World
  Health Organization}.
\bibitem[{Chen et~al.(1990)Chen, Billings \& Grant}]{CHEN1990}
\bibinfo{author}{Chen, S.}, \bibinfo{author}{Billings, S.~A.}, \&
  \bibinfo{author}{Grant, P.~M.} (\bibinfo{year}{1990}).
\newblock \bibinfo{title}{Non-linear system identification using neural
  networks}.
\newblock {\it \bibinfo{journal}{International Journal of Control}\/},  {\it
  \bibinfo{volume}{51}\/}, \bibinfo{pages}{1191--1214}.
  \DOIprefix\doi{10.1080/00207179008934126}.
\bibitem[{Cheng et~al.(2016)Cheng, Wang, Kamalapurkar, Dinh, Bellman \&
  Dixon}]{Cheng2016}
\bibinfo{author}{Cheng, T.-H.}, \bibinfo{author}{Wang, Q.},
  \bibinfo{author}{Kamalapurkar, R.}, \bibinfo{author}{Dinh, H.~T.},
  \bibinfo{author}{Bellman, M.}, \& \bibinfo{author}{Dixon, W.~E.}
  (\bibinfo{year}{2016}).
\newblock \bibinfo{title}{Identification-based closed-loop {NMES} limb tracking
  with amplitude-modulated control input}.
\newblock {\it \bibinfo{journal}{{IEEE} Transactions on Cybernetics}\/},  {\it
  \bibinfo{volume}{46}\/}, \bibinfo{pages}{1679--1690}.
  \DOIprefix\doi{10.1109/tcyb.2015.2453402}.
\bibitem[{Chu et~al.(1990)Chu, Shoureshi \& Tenorio}]{Chu1990}
\bibinfo{author}{Chu, S.}, \bibinfo{author}{Shoureshi, R.}, \&
  \bibinfo{author}{Tenorio, M.} (\bibinfo{year}{1990}).
\newblock \bibinfo{title}{Neural networks for system identification}.
\newblock {\it \bibinfo{journal}{{IEEE} Control Systems Magazine}\/},  {\it
  \bibinfo{volume}{10}\/}, \bibinfo{pages}{31--35}.
  \DOIprefix\doi{10.1109/37.55121}.
\bibitem[{Doucet et~al.(2012)Doucet, Lam \& Griffin}]{doucet}
\bibinfo{author}{Doucet, B.}, \bibinfo{author}{Lam, A.}, \&
  \bibinfo{author}{Griffin, L.} (\bibinfo{year}{2012}).
\newblock \bibinfo{title}{Neuromuscular electrical stimulation for skeletal
  muscle function}.
\newblock {\it \bibinfo{journal}{The Yale journal of biology and medicine}\/},
  {\it \bibinfo{volume}{85}\/}, \bibinfo{pages}{201--215}.
\bibitem[{Downey et~al.(2015)Downey, Cheng, Bellman \& Dixon}]{Downey2015_b}
\bibinfo{author}{Downey, R.~J.}, \bibinfo{author}{Cheng, T.-H.},
  \bibinfo{author}{Bellman, M.~J.}, \& \bibinfo{author}{Dixon, W.~E.}
  (\bibinfo{year}{2015}).
\newblock \bibinfo{title}{Closed-loop asynchronous neuromuscular electrical
  stimulation prolongs functional movements in the lower body}.
\newblock {\it \bibinfo{journal}{{IEEE} Transactions on Neural Systems and
  Rehabilitation Engineering}\/},  {\it \bibinfo{volume}{23}\/},
  \bibinfo{pages}{1117--1127}. \DOIprefix\doi{10.1109/tnsre.2015.2427658}.
\bibitem[{Downey et~al.(2013)Downey, Cheng \& Dixon}]{Downey2013}
\bibinfo{author}{Downey, R.~J.}, \bibinfo{author}{Cheng, T.-H.}, \&
  \bibinfo{author}{Dixon, W.~E.} (\bibinfo{year}{2013}).
\newblock \bibinfo{title}{Tracking control of a human limb during asynchronous
  neuromuscular electrical stimulation}.
\newblock In {\it \bibinfo{booktitle}{52nd {IEEE} Conference on Decision and
  Control}\/}.
\newblock \bibinfo{publisher}{{IEEE}}.
\newblock \DOIprefix\doi{10.1109/cdc.2013.6759872}.
\bibitem[{Ferrarin et~al.(2001)Ferrarin, Palazzo, Riener \&
  Quintern}]{Ferrarin2001}
\bibinfo{author}{Ferrarin, M.}, \bibinfo{author}{Palazzo, F.},
  \bibinfo{author}{Riener, R.}, \& \bibinfo{author}{Quintern, J.}
  (\bibinfo{year}{2001}).
\newblock \bibinfo{title}{Model-based control of {FES}-induced single joint
  movements}.
\newblock {\it \bibinfo{journal}{{IEEE} Transactions on Neural Systems and
  Rehabilitation Engineering}\/},  {\it \bibinfo{volume}{9}\/},
  \bibinfo{pages}{245--257}. \DOIprefix\doi{10.1109/7333.948452}.
\bibitem[{Gaino et~al.(2017)Gaino, Covacic, Teixeira, Cardim,
  Assun{\c{c}}{\~{a}}o, de~Carvalho \& Sanches}]{Gaino2017}
\bibinfo{author}{Gaino, R.}, \bibinfo{author}{Covacic, M.},
  \bibinfo{author}{Teixeira, M.}, \bibinfo{author}{Cardim, R.},
  \bibinfo{author}{Assun{\c{c}}{\~{a}}o, E.}, \bibinfo{author}{de~Carvalho,
  A.}, \& \bibinfo{author}{Sanches, M.} (\bibinfo{year}{2017}).
\newblock \bibinfo{title}{Electrical stimulation tracking control for
  paraplegic patients using t{\textendash}s fuzzy models}.
\newblock {\it \bibinfo{journal}{Fuzzy Sets and Systems}\/},  {\it
  \bibinfo{volume}{314}\/}, \bibinfo{pages}{1--23}.
  \DOIprefix\doi{10.1016/j.fss.2016.06.005}.
\bibitem[{Gaino et~al.(2020)Gaino, Covacic, Cardim, Sanches, Carvalho, Biazeto
  \& Teixeira}]{Gaino2020}
\bibinfo{author}{Gaino, R.}, \bibinfo{author}{Covacic, M.~R.},
  \bibinfo{author}{Cardim, R.}, \bibinfo{author}{Sanches, M. A.~A.},
  \bibinfo{author}{Carvalho, A. A.~D.}, \bibinfo{author}{Biazeto, A.~R.}, \&
  \bibinfo{author}{Teixeira, M. C.~M.} (\bibinfo{year}{2020}).
\newblock \bibinfo{title}{Discrete takagi-sugeno fuzzy models applied to
  control the knee joint movement of paraplegic patients}.
\newblock {\it \bibinfo{journal}{{IEEE} Access}\/},  {\it
  \bibinfo{volume}{8}\/}, \bibinfo{pages}{32714--32726}.
  \DOIprefix\doi{10.1109/access.2020.2971908}.
\bibitem[{Gobbo et~al.(2014)Gobbo, Maffiuletti, Orizio \& Minetto}]{Gobbo2014}
\bibinfo{author}{Gobbo, M.}, \bibinfo{author}{Maffiuletti, N.~A.},
  \bibinfo{author}{Orizio, C.}, \& \bibinfo{author}{Minetto, M.~A.}
  (\bibinfo{year}{2014}).
\newblock \bibinfo{title}{Muscle motor point identification is essential for
  optimizing neuromuscular electrical stimulation use}.
\newblock {\it \bibinfo{journal}{Journal of {NeuroEngineering} and
  Rehabilitation}\/},  {\it \bibinfo{volume}{11}\/}, \bibinfo{pages}{17}.
  \DOIprefix\doi{10.1186/1743-0003-11-17}.
\bibitem[{Gregory et~al.(2007)Gregory, Dixon \& Bickel}]{Gregory2007}
\bibinfo{author}{Gregory, C.~M.}, \bibinfo{author}{Dixon, W.}, \&
  \bibinfo{author}{Bickel, C.~S.} (\bibinfo{year}{2007}).
\newblock \bibinfo{title}{Impact of varying pulse frequency and duration on
  muscle torque production and fatigue}.
\newblock {\it \bibinfo{journal}{Muscle {\&} Nerve}\/},  {\it
  \bibinfo{volume}{35}\/}, \bibinfo{pages}{504--509}.
  \DOIprefix\doi{10.1002/mus.20710}.
\bibitem[{Hmed et~al.(2017)Hmed, Bakir, Garnier, Binczak \&
  Sakly}]{BenHmed2017}
\bibinfo{author}{Hmed, A.~B.}, \bibinfo{author}{Bakir, T.},
  \bibinfo{author}{Garnier, Y.}, \bibinfo{author}{Binczak, S.}, \&
  \bibinfo{author}{Sakly, A.} (\bibinfo{year}{2017}).
\newblock \bibinfo{title}{A novel strategy for adjusting current pulse
  amplitude of {FES}-systems with {PID} based on {PSO} algorithm method to
  control the muscle force}.
\newblock In {\it \bibinfo{booktitle}{Proceedings of the 14th International
  Conference on Informatics in Control, Automation and Robotics}\/}.
\newblock \bibinfo{publisher}{{SCITEPRESS} - Science and Technology
  Publications}.
\newblock \DOIprefix\doi{10.5220/0006473306640669}.
\bibitem[{Ho et~al.(2014)Ho, Triolo, Elias, Kilgore, DiMarco, Bogie, Vette,
  Audu, Kobetic, Chang, Chan, Dukelow, Bourbeau, Brose, Gustafson, Kiss \&
  Mushahwar}]{Ho2014}
\bibinfo{author}{Ho, C.~H.}, \bibinfo{author}{Triolo, R.~J.},
  \bibinfo{author}{Elias, A.~L.}, \bibinfo{author}{Kilgore, K.~L.},
  \bibinfo{author}{DiMarco, A.~F.}, \bibinfo{author}{Bogie, K.},
  \bibinfo{author}{Vette, A.~H.}, \bibinfo{author}{Audu, M.~L.},
  \bibinfo{author}{Kobetic, R.}, \bibinfo{author}{Chang, S.~R.},
  \bibinfo{author}{Chan, K.~M.}, \bibinfo{author}{Dukelow, S.},
  \bibinfo{author}{Bourbeau, D.~J.}, \bibinfo{author}{Brose, S.~W.},
  \bibinfo{author}{Gustafson, K.~J.}, \bibinfo{author}{Kiss, Z.~H.}, \&
  \bibinfo{author}{Mushahwar, V.~K.} (\bibinfo{year}{2014}).
\newblock \bibinfo{title}{Functional electrical stimulation and spinal cord
  injury}.
\newblock {\it \bibinfo{journal}{Physical Medicine and Rehabilitation Clinics
  of North America}\/},  {\it \bibinfo{volume}{25}\/},
  \bibinfo{pages}{631--654}. \DOIprefix\doi{10.1016/j.pmr.2014.05.001}.
\bibitem[{Hornik et~al.(1989)Hornik, Stinchcombe \&
  White}]{hornik1989multilayer}
\bibinfo{author}{Hornik, K.}, \bibinfo{author}{Stinchcombe, M.}, \&
  \bibinfo{author}{White, H.} (\bibinfo{year}{1989}).
\newblock \bibinfo{title}{Multilayer feedforward networks are universal
  approximators}.
\newblock {\it \bibinfo{journal}{Neural Networks}\/},  {\it
  \bibinfo{volume}{2}\/}, \bibinfo{pages}{359--366}.
\bibitem[{Jezernik et~al.(2004)Jezernik, Wassink \& Keller}]{Jezernik2004}
\bibinfo{author}{Jezernik, S.}, \bibinfo{author}{Wassink, R.}, \&
  \bibinfo{author}{Keller, T.} (\bibinfo{year}{2004}).
\newblock \bibinfo{title}{Sliding mode closed-loop control of {FES}:
  Controlling the shank movement}.
\newblock {\it \bibinfo{journal}{{IEEE} Transactions on Biomedical
  Engineering}\/},  {\it \bibinfo{volume}{51}\/}, \bibinfo{pages}{263--272}.
  \DOIprefix\doi{10.1109/tbme.2003.820393}.
\bibitem[{Kapadia et~al.(2020)Kapadia, Moineau \&
  Popovic}]{kapadia2020functional}
\bibinfo{author}{Kapadia, N.}, \bibinfo{author}{Moineau, B.}, \&
  \bibinfo{author}{Popovic, M.~R.} (\bibinfo{year}{2020}).
\newblock \bibinfo{title}{Functional electrical stimulation therapy for
  retraining reaching and grasping after spinal cord injury and stroke}.
\newblock {\it \bibinfo{journal}{Frontiers in Neuroscience}\/},  {\it
  \bibinfo{volume}{14}\/}, \bibinfo{pages}{718}.
\bibitem[{Kawai et~al.(2014)Kawai, Downey, Kawai \& Dixon}]{Kawai2014}
\bibinfo{author}{Kawai, Y.}, \bibinfo{author}{Downey, R.~J.},
  \bibinfo{author}{Kawai, H.}, \& \bibinfo{author}{Dixon, W.~E.}
  (\bibinfo{year}{2014}).
\newblock \bibinfo{title}{Co-contraction of antagonist bi-articular muscles for
  tracking control of human limb}.
\newblock In {\it \bibinfo{booktitle}{2014 American Control Conference}\/}.
\newblock \bibinfo{publisher}{{IEEE}}.
\newblock \DOIprefix\doi{10.1109/acc.2014.6859102}.
\bibitem[{Khadra et~al.(2016)Khadra, Qudeiri \& Alkahtani}]{AbuKhadra2016}
\bibinfo{author}{Khadra, F.~A.}, \bibinfo{author}{Qudeiri, J.~A.}, \&
  \bibinfo{author}{Alkahtani, M.} (\bibinfo{year}{2016}).
\newblock \bibinfo{title}{Optimization of the parameters of {RISE} feedback
  controller using genetic algorithm}.
\newblock {\it \bibinfo{journal}{Mathematical Problems in Engineering}\/},
  {\it \bibinfo{volume}{2016}\/}, \bibinfo{pages}{1--9}.
  \DOIprefix\doi{10.1155/2016/3863147}.
\bibitem[{Kushima et~al.(2015)Kushima, Kawataka, Kawai, Kawai \&
  Dixon}]{Kushima2015}
\bibinfo{author}{Kushima, Y.}, \bibinfo{author}{Kawataka, K.},
  \bibinfo{author}{Kawai, H.}, \bibinfo{author}{Kawai, Y.}, \&
  \bibinfo{author}{Dixon, W.~E.} (\bibinfo{year}{2015}).
\newblock \bibinfo{title}{{FES} knee bending and stretching system with
  {RISE}-based tracking control for human limb}.
\newblock In {\it \bibinfo{booktitle}{2015 {IEEE} Conference on Control
  Applications ({CCA})}\/}.
\newblock \bibinfo{publisher}{{IEEE}}.
\newblock \DOIprefix\doi{10.1109/cca.2015.7320727}.
\bibitem[{Laubacher et~al.(2017)Laubacher, Aks\"{o}z, Riener, Binder-Macleod \&
  Hunt}]{Laubacher2017}
\bibinfo{author}{Laubacher, M.}, \bibinfo{author}{Aks\"{o}z, A.~E.},
  \bibinfo{author}{Riener, R.}, \bibinfo{author}{Binder-Macleod, S.}, \&
  \bibinfo{author}{Hunt, K.~J.} (\bibinfo{year}{2017}).
\newblock \bibinfo{title}{Power output and fatigue properties using spatially
  distributed sequential stimulation in a dynamic knee extension task}.
\newblock {\it \bibinfo{journal}{European Journal of Applied Physiology}\/},
  {\it \bibinfo{volume}{117}\/}, \bibinfo{pages}{1787--1798}.
  \DOIprefix\doi{10.1007/s00421-017-3675-0}.
\bibitem[{Lew et~al.(2016)Lew, Alavi, Randhawa \& Menon}]{Lew2016}
\bibinfo{author}{Lew, B.}, \bibinfo{author}{Alavi, N.},
  \bibinfo{author}{Randhawa, B.~K.}, \& \bibinfo{author}{Menon, C.}
  (\bibinfo{year}{2016}).
\newblock \bibinfo{title}{An exploratory investigation on the use of
  closed-loop electrical stimulation to assist individuals with stroke to
  perform fine movements with their hemiparetic arm}.
\newblock {\it \bibinfo{journal}{Frontiers in Bioengineering and
  Biotechnology}\/},  {\it \bibinfo{volume}{4}\/}.
  \DOIprefix\doi{10.3389/fbioe.2016.00020}.
\bibitem[{Lynch \& Popovic(2008)}]{LynchFES}
\bibinfo{author}{Lynch, C.~L.}, \& \bibinfo{author}{Popovic, M.~R.}
  (\bibinfo{year}{2008}).
\newblock \bibinfo{title}{Functional electrical stimulation}.
\newblock {\it \bibinfo{journal}{{IEEE} Control Systems}\/},  {\it
  \bibinfo{volume}{28}\/}, \bibinfo{pages}{40--50}.
  \DOIprefix\doi{10.1109/mcs.2007.914689}.
\bibitem[{Lynch \& Popovic(2012)}]{Lynch2012}
\bibinfo{author}{Lynch, C.~L.}, \& \bibinfo{author}{Popovic, M.~R.}
  (\bibinfo{year}{2012}).
\newblock \bibinfo{title}{A comparison of closed-loop control algorithms for
  regulating electrically stimulated knee movements in individuals with spinal
  cord injury}.
\newblock {\it \bibinfo{journal}{{IEEE} Transactions on Neural Systems and
  Rehabilitation Engineering}\/},  {\it \bibinfo{volume}{20}\/},
  \bibinfo{pages}{539--548}. \DOIprefix\doi{10.1109/tnsre.2012.2185065}.
\bibitem[{Maffiuletti(2010)}]{Maffiuletti2010}
\bibinfo{author}{Maffiuletti, N.~A.} (\bibinfo{year}{2010}).
\newblock \bibinfo{title}{Physiological and methodological considerations for
  the use of neuromuscular electrical stimulation}.
\newblock {\it \bibinfo{journal}{European Journal of Applied Physiology}\/},
  {\it \bibinfo{volume}{110}\/}, \bibinfo{pages}{223--234}.
  \DOIprefix\doi{10.1007/s00421-010-1502-y}.
\bibitem[{{Makkar} et~al.(2007){Makkar}, {Hu}, {Sawyer} \&
  {Dixon}}]{Makkar2007}
\bibinfo{author}{{Makkar}, C.}, \bibinfo{author}{{Hu}, G.},
  \bibinfo{author}{{Sawyer}, W.~G.}, \& \bibinfo{author}{{Dixon}, W.~E.}
  (\bibinfo{year}{2007}).
\newblock \bibinfo{title}{Lyapunov-based tracking control in the presence of
  uncertain nonlinear parameterizable friction}.
\newblock {\it \bibinfo{journal}{IEEE Transactions on Automatic Control}\/},
  {\it \bibinfo{volume}{52}\/}, \bibinfo{pages}{1988--1994}.
  \DOIprefix\doi{10.1109/TAC.2007.904254}.
\bibitem[{Marquez-Chin \& Popovic(2020)}]{marquez2020functional}
\bibinfo{author}{Marquez-Chin, C.}, \& \bibinfo{author}{Popovic, M.~R.}
  (\bibinfo{year}{2020}).
\newblock \bibinfo{title}{Functional electrical stimulation therapy for
  restoration of motor function after spinal cord injury and stroke: a review}.
\newblock {\it \bibinfo{journal}{BioMedical Engineering OnLine}\/},  {\it
  \bibinfo{volume}{19}\/}, \bibinfo{pages}{1--25}.
\bibitem[{Mohammed et~al.(2012)Mohammed, Poignet, Fraisse \&
  Guiraud}]{Mohammed2012}
\bibinfo{author}{Mohammed, S.}, \bibinfo{author}{Poignet, P.},
  \bibinfo{author}{Fraisse, P.}, \& \bibinfo{author}{Guiraud, D.}
  (\bibinfo{year}{2012}).
\newblock \bibinfo{title}{Toward lower limbs movement restoration with
  input{\textendash}output feedback linearization and model predictive control
  through functional electrical stimulation}.
\newblock {\it \bibinfo{journal}{Control Engineering Practice}\/},  {\it
  \bibinfo{volume}{20}\/}, \bibinfo{pages}{182--195}.
  \DOIprefix\doi{10.1016/j.conengprac.2011.10.010}.
\bibitem[{M\"{u}ller et~al.(2017)M\"{u}ller, Balligand, Seel \&
  Schauer}]{Mller2017}
\bibinfo{author}{M\"{u}ller, P.}, \bibinfo{author}{Balligand, C.},
  \bibinfo{author}{Seel, T.}, \& \bibinfo{author}{Schauer, T.}
  (\bibinfo{year}{2017}).
\newblock \bibinfo{title}{Iterative learning control and system identification
  of the antagonistic knee muscle complex during gait using functional
  electrical stimulation}.
\newblock {\it \bibinfo{journal}{{IFAC}-{PapersOnLine}}\/},  {\it
  \bibinfo{volume}{50}\/}, \bibinfo{pages}{8786--8791}.
  \DOIprefix\doi{10.1016/j.ifacol.2017.08.1738}.
\bibitem[{Narendra \& Parthasarathy(1990)}]{Narendra1990}
\bibinfo{author}{Narendra, K.}, \& \bibinfo{author}{Parthasarathy, K.}
  (\bibinfo{year}{1990}).
\newblock \bibinfo{title}{Identification and control of dynamical systems using
  neural networks}.
\newblock {\it \bibinfo{journal}{{IEEE} Transactions on Neural Networks}\/},
  {\it \bibinfo{volume}{1}\/}, \bibinfo{pages}{4--27}.
  \DOIprefix\doi{10.1109/72.80202}.
\bibitem[{Nunes et~al.(2019)Nunes, Teodoro, Sanches, de~Araujo, Teixeira \&
  Carvalho}]{nunes2018}
\bibinfo{author}{Nunes, W. R. B.~M.}, \bibinfo{author}{Teodoro, R.~G.},
  \bibinfo{author}{Sanches, M. A.~A.}, \bibinfo{author}{de~Araujo, R.~A.},
  \bibinfo{author}{Teixeira, M. C.~M.}, \& \bibinfo{author}{Carvalho, A.~A.}
  (\bibinfo{year}{2019}).
\newblock \bibinfo{title}{Switched controller applied to functional electrical
  stimulation of lower limbs under fatigue conditions: A linear analysis}.
\newblock In {\it \bibinfo{booktitle}{{XXVI} Brazilian Congress on Biomedical
  Engineering}\/} (pp. \bibinfo{pages}{383--390}).
\newblock \bibinfo{publisher}{Springer Singapore}.
\newblock \DOIprefix\doi{10.1007/978-981-13-2119-1_59}.
\bibitem[{Page \& Freeman(2020)}]{Page2020}
\bibinfo{author}{Page, A.}, \& \bibinfo{author}{Freeman, C.}
  (\bibinfo{year}{2020}).
\newblock \bibinfo{title}{Point-to-point repetitive control of functional
  electrical stimulation for drop-foot}.
\newblock {\it \bibinfo{journal}{Control Engineering Practice}\/},  {\it
  \bibinfo{volume}{96}\/}, \bibinfo{pages}{104280}.
  \DOIprefix\doi{10.1016/j.conengprac.2019.104280}.
\bibitem[{{Patre} et~al.(2008){Patre}, {MacKunis}, {Makkar} \&
  {Dixon}}]{Patre2008}
\bibinfo{author}{{Patre}, P.~M.}, \bibinfo{author}{{MacKunis}, W.},
  \bibinfo{author}{{Makkar}, C.}, \& \bibinfo{author}{{Dixon}, W.~E.}
  (\bibinfo{year}{2008}).
\newblock \bibinfo{title}{Asymptotic tracking for systems with structured and
  unstructured uncertainties}.
\newblock {\it \bibinfo{journal}{IEEE Transactions on Control Systems
  Technology}\/},  {\it \bibinfo{volume}{16}\/}, \bibinfo{pages}{373--379}.
  \DOIprefix\doi{10.1109/TCST.2007.908227}.
\bibitem[{Peckham \& Knutson(2005)}]{Peckham2005}
\bibinfo{author}{Peckham, P.~H.}, \& \bibinfo{author}{Knutson, J.~S.}
  (\bibinfo{year}{2005}).
\newblock \bibinfo{title}{Functional electrical stimulation for neuromuscular
  applications}.
\newblock {\it \bibinfo{journal}{Annual Review of Biomedical Engineering}\/},
  {\it \bibinfo{volume}{7}\/}, \bibinfo{pages}{327--360}.
  \DOIprefix\doi{10.1146/annurev.bioeng.6.040803.140103}.
\bibitem[{Popovi{\'{c}}(2014)}]{Popovi2014}
\bibinfo{author}{Popovi{\'{c}}, D.~B.} (\bibinfo{year}{2014}).
\newblock \bibinfo{title}{Advances in functional electrical stimulation
  ({FES})}.
\newblock {\it \bibinfo{journal}{Journal of Electromyography and
  Kinesiology}\/},  {\it \bibinfo{volume}{24}\/}, \bibinfo{pages}{795--802}.
  \DOIprefix\doi{10.1016/j.jelekin.2014.09.008}.
\bibitem[{Previdi(2002)}]{Previdi2002}
\bibinfo{author}{Previdi, F.} (\bibinfo{year}{2002}).
\newblock \bibinfo{title}{Identification of black-box nonlinear models for
  lower limb movement control using functional electrical stimulation}.
\newblock {\it \bibinfo{journal}{Control Engineering Practice}\/},  {\it
  \bibinfo{volume}{10}\/}, \bibinfo{pages}{91--99}.
  \DOIprefix\doi{10.1016/s0967-0661(01)00128-9}.
\bibitem[{Previdi \& Carpanzano(2003)}]{Previdi2003}
\bibinfo{author}{Previdi, F.}, \& \bibinfo{author}{Carpanzano, E.}
  (\bibinfo{year}{2003}).
\newblock \bibinfo{title}{Design of a gain scheduling controller for knee-joint
  angle control by using functional electrical stimulation}.
\newblock {\it \bibinfo{journal}{{IEEE} Transactions on Control Systems
  Technology}\/},  {\it \bibinfo{volume}{11}\/}, \bibinfo{pages}{310--324}.
  \DOIprefix\doi{10.1109/tcst.2003.810380}.
\bibitem[{dos Santos et~al.(2015)dos Santos, Gaino, Covacic, Teixeira,
  de~Carvalho, Assun{\c{c}}{\~{a}}o, Cardim \& Sanches}]{Santos2015}
\bibinfo{author}{dos Santos, N.~M.}, \bibinfo{author}{Gaino, R.},
  \bibinfo{author}{Covacic, M.~R.}, \bibinfo{author}{Teixeira, M. C.~M.},
  \bibinfo{author}{de~Carvalho, A.~A.}, \bibinfo{author}{Assun{\c{c}}{\~{a}}o,
  E.}, \bibinfo{author}{Cardim, R.}, \& \bibinfo{author}{Sanches, M. A.~A.}
  (\bibinfo{year}{2015}).
\newblock \bibinfo{title}{Robust control of the knee joint angle of paraplegic
  patients considering norm-bounded uncertainties}.
\newblock {\it \bibinfo{journal}{Mathematical Problems in Engineering}\/},
  {\it \bibinfo{volume}{2015}\/}, \bibinfo{pages}{1--8}.
  \DOIprefix\doi{10.1155/2015/736246}.
\bibitem[{Sharma et~al.(2012)Sharma, Gregory, Johnson \& Dixon}]{Sharma2012}
\bibinfo{author}{Sharma, N.}, \bibinfo{author}{Gregory, C.~M.},
  \bibinfo{author}{Johnson, M.}, \& \bibinfo{author}{Dixon, W.~E.}
  (\bibinfo{year}{2012}).
\newblock \bibinfo{title}{Closed-loop neural network-based {NMES} control for
  human limb tracking}.
\newblock {\it \bibinfo{journal}{{IEEE} Transactions on Control Systems
  Technology}\/},  {\it \bibinfo{volume}{20}\/}, \bibinfo{pages}{712--725}.
  \DOIprefix\doi{10.1109/tcst.2011.2125792}.
\bibitem[{Sharma et~al.(2009)Sharma, Stegath, Gregory \& Dixon}]{Sharma2009}
\bibinfo{author}{Sharma, N.}, \bibinfo{author}{Stegath, K.},
  \bibinfo{author}{Gregory, C.}, \& \bibinfo{author}{Dixon, W.}
  (\bibinfo{year}{2009}).
\newblock \bibinfo{title}{Nonlinear neuromuscular electrical stimulation
  tracking control of a human limb}.
\newblock {\it \bibinfo{journal}{{IEEE} Transactions on Neural Systems and
  Rehabilitation Engineering}\/},  {\it \bibinfo{volume}{17}\/},
  \bibinfo{pages}{576--584}. \DOIprefix\doi{10.1109/tnsre.2009.2023294}.
\bibitem[{Stegath et~al.(2007)Stegath, Sharma, Gregory \& Dixon}]{Stegath2007}
\bibinfo{author}{Stegath, K.}, \bibinfo{author}{Sharma, N.},
  \bibinfo{author}{Gregory, C.~M.}, \& \bibinfo{author}{Dixon, W.~E.}
  (\bibinfo{year}{2007}).
\newblock \bibinfo{title}{Experimental demonstration of {RISE}-based {NMES} of
  human quadriceps muscle}.
\newblock In {\it \bibinfo{booktitle}{2007 {IEEE}/{NIH} Life Science Systems
  and Applications Workshop}\/}.
\newblock \bibinfo{publisher}{{IEEE}}.
\newblock \DOIprefix\doi{10.1109/lssa.2007.4400880}.
\bibitem[{Stegath et~al.(2008)Stegath, Sharma, Gregory \& Dixon}]{Stegath2008}
\bibinfo{author}{Stegath, K.}, \bibinfo{author}{Sharma, N.},
  \bibinfo{author}{Gregory, C.~M.}, \& \bibinfo{author}{Dixon, W.~E.}
  (\bibinfo{year}{2008}).
\newblock \bibinfo{title}{Nonlinear tracking control of a human limb via
  neuromuscular electrical stimulation}.
\newblock In {\it \bibinfo{booktitle}{2008 American Control Conference}\/}.
\newblock \bibinfo{publisher}{{IEEE}}.
\newblock \DOIprefix\doi{10.1109/acc.2008.4586776}.
\bibitem[{Teodoro et~al.(2020)Teodoro, Nunes, de~Araujo, Sanches, Teixeira \&
  de~Carvalho}]{Teodoro2020}
\bibinfo{author}{Teodoro, R.~G.}, \bibinfo{author}{Nunes, W.~R.},
  \bibinfo{author}{de~Araujo, R.~A.}, \bibinfo{author}{Sanches, M.~A.},
  \bibinfo{author}{Teixeira, M.~C.}, \& \bibinfo{author}{de~Carvalho, A.~A.}
  (\bibinfo{year}{2020}).
\newblock \bibinfo{title}{Robust switched control design for electrically
  stimulated lower limbs: A linear model analysis in healthy and spinal cord
  injured subjects}.
\newblock {\it \bibinfo{journal}{Control Engineering Practice}\/},  {\it
  \bibinfo{volume}{102}\/}, \bibinfo{pages}{104530}.
  \DOIprefix\doi{10.1016/j.conengprac.2020.104530}.
\bibitem[{Utkin(2013)}]{utkin2013sliding}
\bibinfo{author}{Utkin, V.~I.} (\bibinfo{year}{2013}).
\newblock {\it \bibinfo{title}{Sliding modes in control and optimization}\/}.
\newblock \bibinfo{publisher}{Springer, Berlin, Heidelberg}.
\newblock \DOIprefix\doi{10.1007/978-3-642-84379-2}.
\bibitem[{Wagner et~al.(2018)Wagner, Mignardot, Goff-Mignardot, Demesmaeker,
  Komi, Capogrosso, Rowald, Se{\'{a}}{\~{n}}ez, Caban, Pirondini, Vat,
  McCracken, Heimgartner, Fodor, Watrin, Seguin, Paoles, Keybus, Eberle,
  Schurch, Pralong, Becce, Prior, Buse, Buschman, Neufeld, Kuster, Carda, von
  Zitzewitz, Delattre, Denison, Lambert, Minassian, Bloch \&
  Courtine}]{Wagner2018}
\bibinfo{author}{Wagner, F.~B.}, \bibinfo{author}{Mignardot, J.-B.},
  \bibinfo{author}{Goff-Mignardot, C. G.~L.}, \bibinfo{author}{Demesmaeker,
  R.}, \bibinfo{author}{Komi, S.}, \bibinfo{author}{Capogrosso, M.},
  \bibinfo{author}{Rowald, A.}, \bibinfo{author}{Se{\'{a}}{\~{n}}ez, I.},
  \bibinfo{author}{Caban, M.}, \bibinfo{author}{Pirondini, E.},
  \bibinfo{author}{Vat, M.}, \bibinfo{author}{McCracken, L.~A.},
  \bibinfo{author}{Heimgartner, R.}, \bibinfo{author}{Fodor, I.},
  \bibinfo{author}{Watrin, A.}, \bibinfo{author}{Seguin, P.},
  \bibinfo{author}{Paoles, E.}, \bibinfo{author}{Keybus, K. V.~D.},
  \bibinfo{author}{Eberle, G.}, \bibinfo{author}{Schurch, B.},
  \bibinfo{author}{Pralong, E.}, \bibinfo{author}{Becce, F.},
  \bibinfo{author}{Prior, J.}, \bibinfo{author}{Buse, N.},
  \bibinfo{author}{Buschman, R.}, \bibinfo{author}{Neufeld, E.},
  \bibinfo{author}{Kuster, N.}, \bibinfo{author}{Carda, S.},
  \bibinfo{author}{von Zitzewitz, J.}, \bibinfo{author}{Delattre, V.},
  \bibinfo{author}{Denison, T.}, \bibinfo{author}{Lambert, H.},
  \bibinfo{author}{Minassian, K.}, \bibinfo{author}{Bloch, J.}, \&
  \bibinfo{author}{Courtine, G.} (\bibinfo{year}{2018}).
\newblock \bibinfo{title}{Targeted neurotechnology restores walking in humans
  with spinal cord injury}.
\newblock {\it \bibinfo{journal}{Nature}\/},  {\it \bibinfo{volume}{563}\/},
  \bibinfo{pages}{65--71}. \DOIprefix\doi{10.1038/s41586-018-0649-2}.
\bibitem[{Wu et~al.(2017)Wu, Wu, Zhang \& Xiong}]{Wu2017}
\bibinfo{author}{Wu, L.}, \bibinfo{author}{Wu, Q.}, \bibinfo{author}{Zhang,
  Q.}, \& \bibinfo{author}{Xiong, C.} (\bibinfo{year}{2017}).
\newblock \bibinfo{title}{Electrically induced joint movement control with
  iterative learning algorithm}.
\newblock In {\it \bibinfo{booktitle}{2017 2nd International Conference on
  Advanced Robotics and Mechatronics ({ICARM})}\/}.
\newblock \bibinfo{publisher}{{IEEE}}.
\newblock \DOIprefix\doi{10.1109/icarm.2017.8273200}.
\bibitem[{Xian et~al.(2003)Xian, Dawson, de~Queiroz \& Chen}]{Xian2003}
\bibinfo{author}{Xian, B.}, \bibinfo{author}{Dawson, D.},
  \bibinfo{author}{de~Queiroz, M.}, \& \bibinfo{author}{Chen, J.}
  (\bibinfo{year}{2003}).
\newblock \bibinfo{title}{A continuous asymptotic tracking control strategy for
  uncertain multi-input nonlinear systems}.
\newblock In {\it \bibinfo{booktitle}{Proceedings of the 2003 {IEEE}
  International Symposium on Intelligent Control {ISIC}-03}\/}.
\newblock \bibinfo{publisher}{{IEEE}}.
\newblock \DOIprefix\doi{10.1109/isic.2003.1253913}.
\bibitem[{{Xian} et~al.(2004){Xian}, {Dawson}, {de Queiroz} \&
  {Chen}}]{Xian2004}
\bibinfo{author}{{Xian}, B.}, \bibinfo{author}{{Dawson}, D.~M.},
  \bibinfo{author}{{de Queiroz}, M.~S.}, \& \bibinfo{author}{{Chen}, J.}
  (\bibinfo{year}{2004}).
\newblock \bibinfo{title}{A continuous asymptotic tracking control strategy for
  uncertain nonlinear systems}.
\newblock {\it \bibinfo{journal}{IEEE Transactions on Automatic Control}\/},
  {\it \bibinfo{volume}{49}\/}, \bibinfo{pages}{1206--1211}.
  \DOIprefix\doi{10.1109/TAC.2004.831148}.
\bibitem[{Yu et~al.(2015)Yu, Huang, Chen, Pan \& Guo}]{Yu2015}
\bibinfo{author}{Yu, H.}, \bibinfo{author}{Huang, S.}, \bibinfo{author}{Chen,
  G.}, \bibinfo{author}{Pan, Y.}, \& \bibinfo{author}{Guo, Z.}
  (\bibinfo{year}{2015}).
\newblock \bibinfo{title}{Human{\textendash}robot interaction control of
  rehabilitation robots with series elastic actuators}.
\newblock {\it \bibinfo{journal}{{IEEE} Transactions on Robotics}\/},  {\it
  \bibinfo{volume}{31}\/}, \bibinfo{pages}{1089--1100}.
  \DOIprefix\doi{10.1109/tro.2015.2457314}.
\bibitem[{Yu et~al.(2013)Yu, Huang, Chen \& Thakor}]{Yu2013}
\bibinfo{author}{Yu, H.}, \bibinfo{author}{Huang, S.}, \bibinfo{author}{Chen,
  G.}, \& \bibinfo{author}{Thakor, N.} (\bibinfo{year}{2013}).
\newblock \bibinfo{title}{Control design of a novel compliant actuator for
  rehabilitation robots}.
\newblock {\it \bibinfo{journal}{Mechatronics}\/},  {\it
  \bibinfo{volume}{23}\/}, \bibinfo{pages}{1072--1083}.
  \DOIprefix\doi{10.1016/j.mechatronics.2013.08.004}.

\end{thebibliography}

\appendix

\section{Supplementary results}\label{app:sup_results} 

\begin{figure}[htb]
	\centering
	\includegraphics[width=1\linewidth]{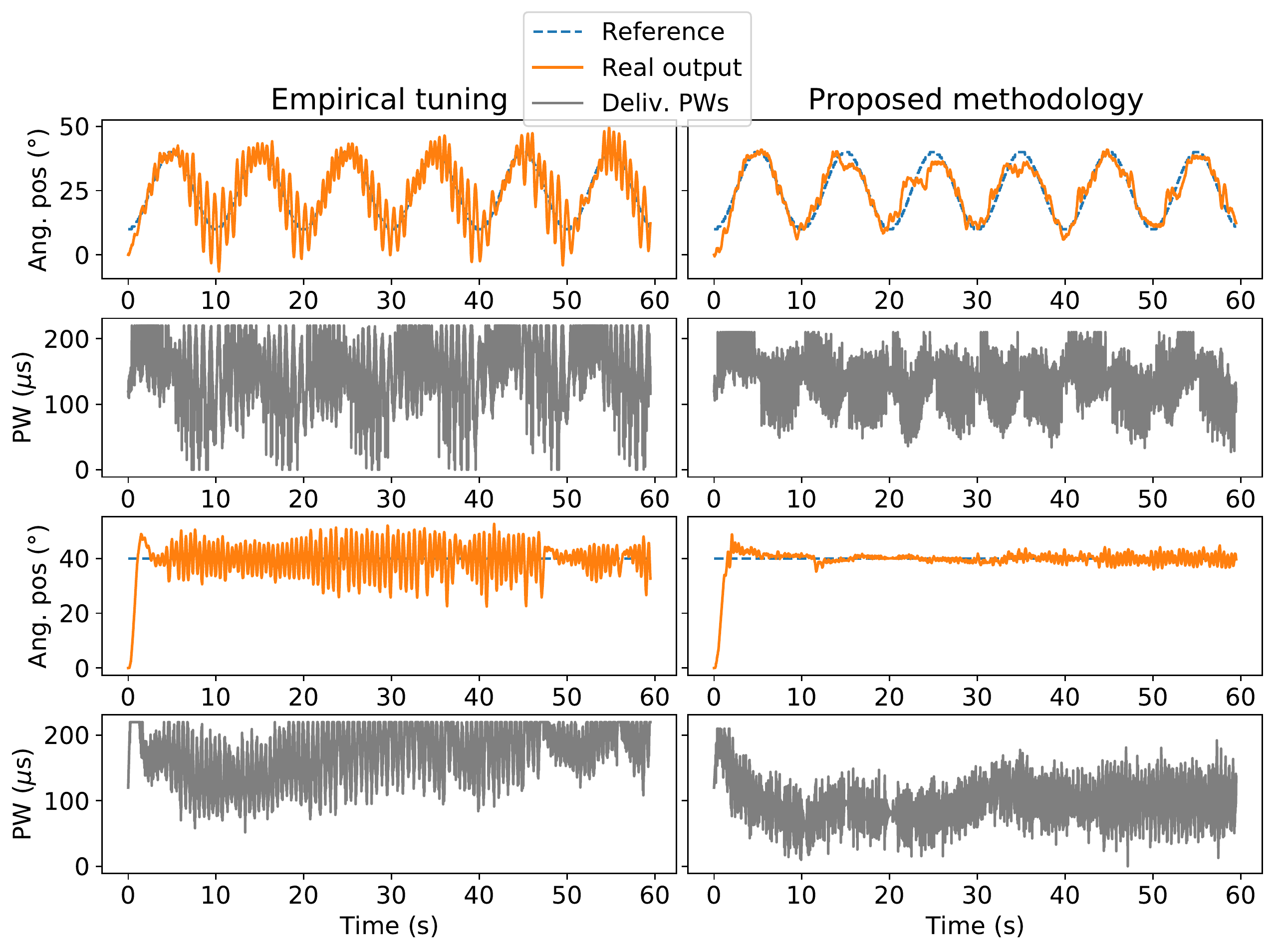}
	\caption{Experimental results for individual H1 comparing empirical gains and the proposed methodology. The first and second rows illustrate the tracking results for the sine wave and the corresponding delivered PWs, respectively. Similarly, the third and fourth rows illustrate the tracking results for the step wave and the corresponding delivered PWs, respectively.}
	\label{fig:h1_results}
\end{figure}

\begin{figure}[htb]
	\centering
	\includegraphics[width=1\linewidth]{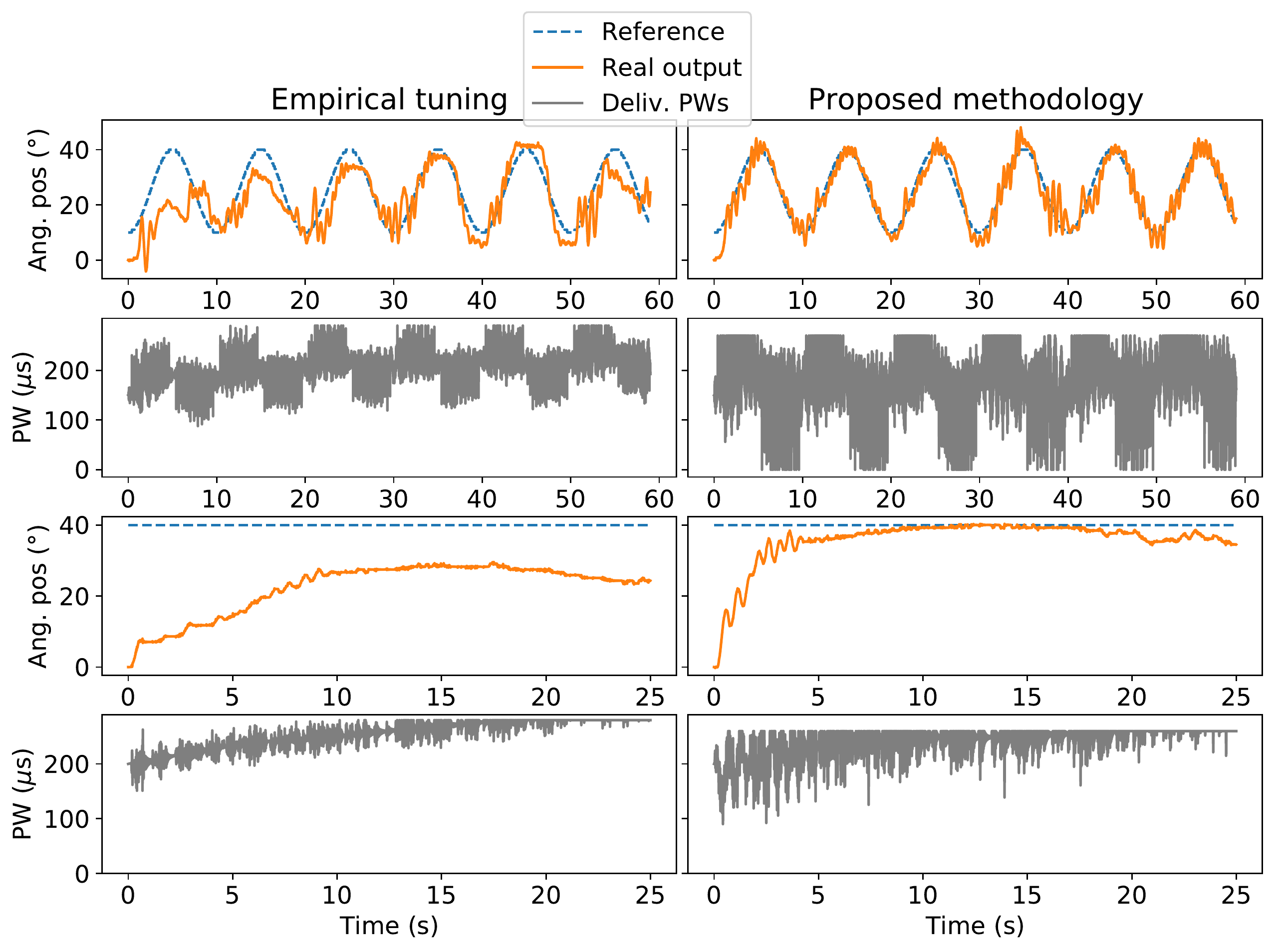}
	\caption{Experimental results for individual H2 comparing empirical gains and the proposed methodology. The first and second rows illustrate the tracking results for the sine wave and the corresponding delivered PWs, respectively. Similarly, the third and fourth rows illustrate the tracking results for the step wave and the corresponding delivered PWs, respectively.}
	\label{fig:h2_results}
\end{figure}

\begin{figure}[htb]
	\centering
	\includegraphics[width=1\linewidth]{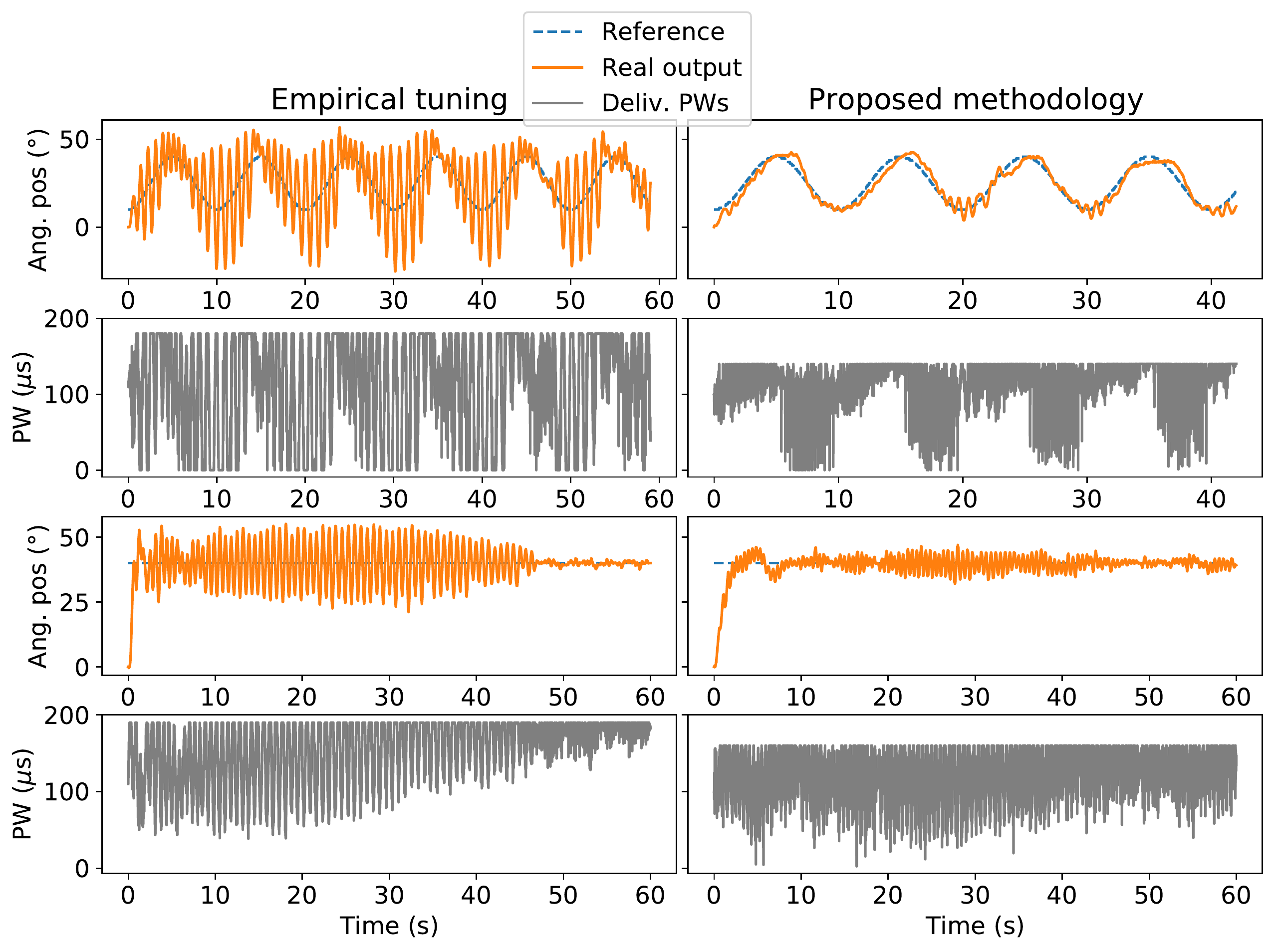}
	\caption{Experimental results for individual H4 comparing empirical gains and the proposed methodology. The first and second rows illustrate the tracking results for the sine wave and the corresponding delivered PWs, respectively. Similarly, the third and fourth rows illustrate the tracking results for the step wave and the corresponding delivered PWs, respectively.}
	\label{fig:h4_results}
\end{figure}

\begin{figure}[!h]
	\centering
	\includegraphics[width=1\linewidth]{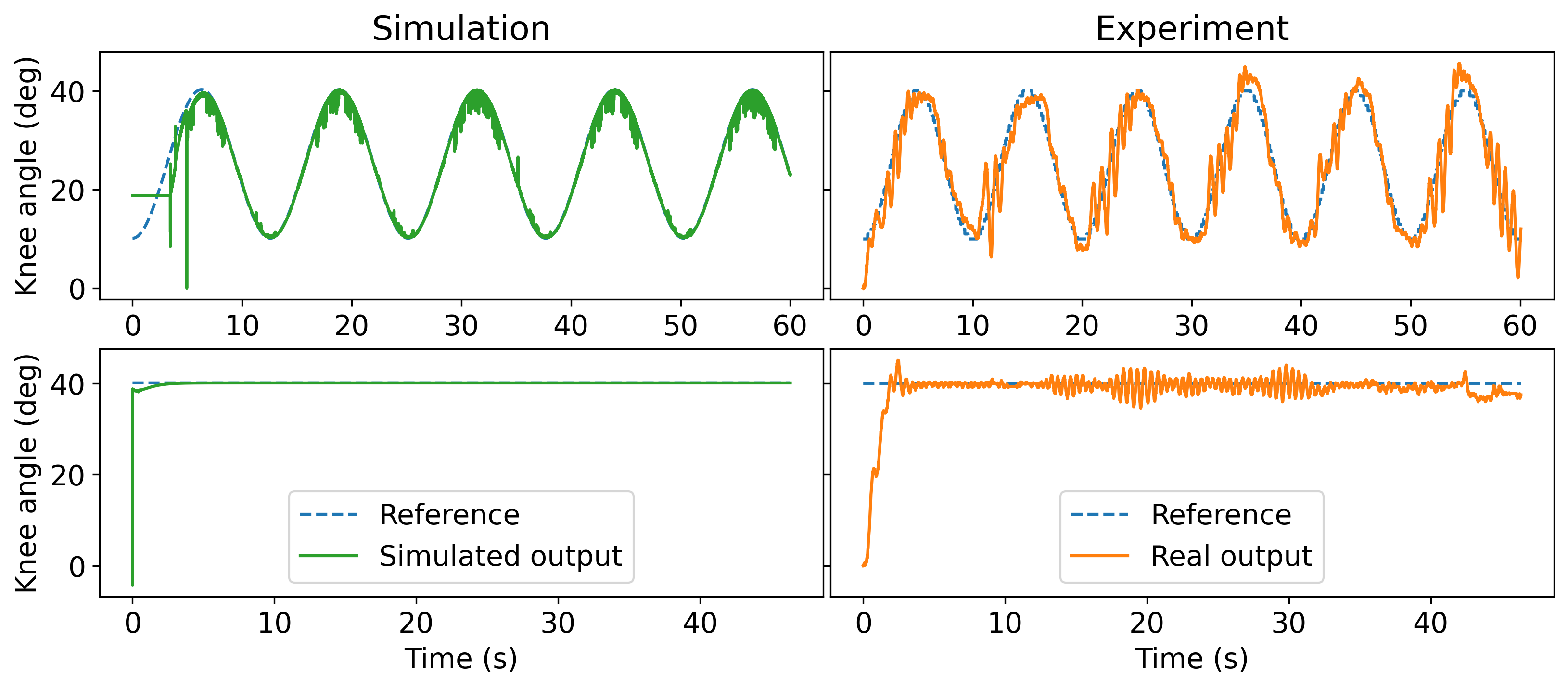}
	\caption{Comparison of simulation and real experiments for individual H2 using past rehabilitation data to identify the nonlinear model.}
	\label{fig:H2_simVSreal}
\end{figure}

\begin{figure}[!h]
	\centering
	\includegraphics[width=1\linewidth]{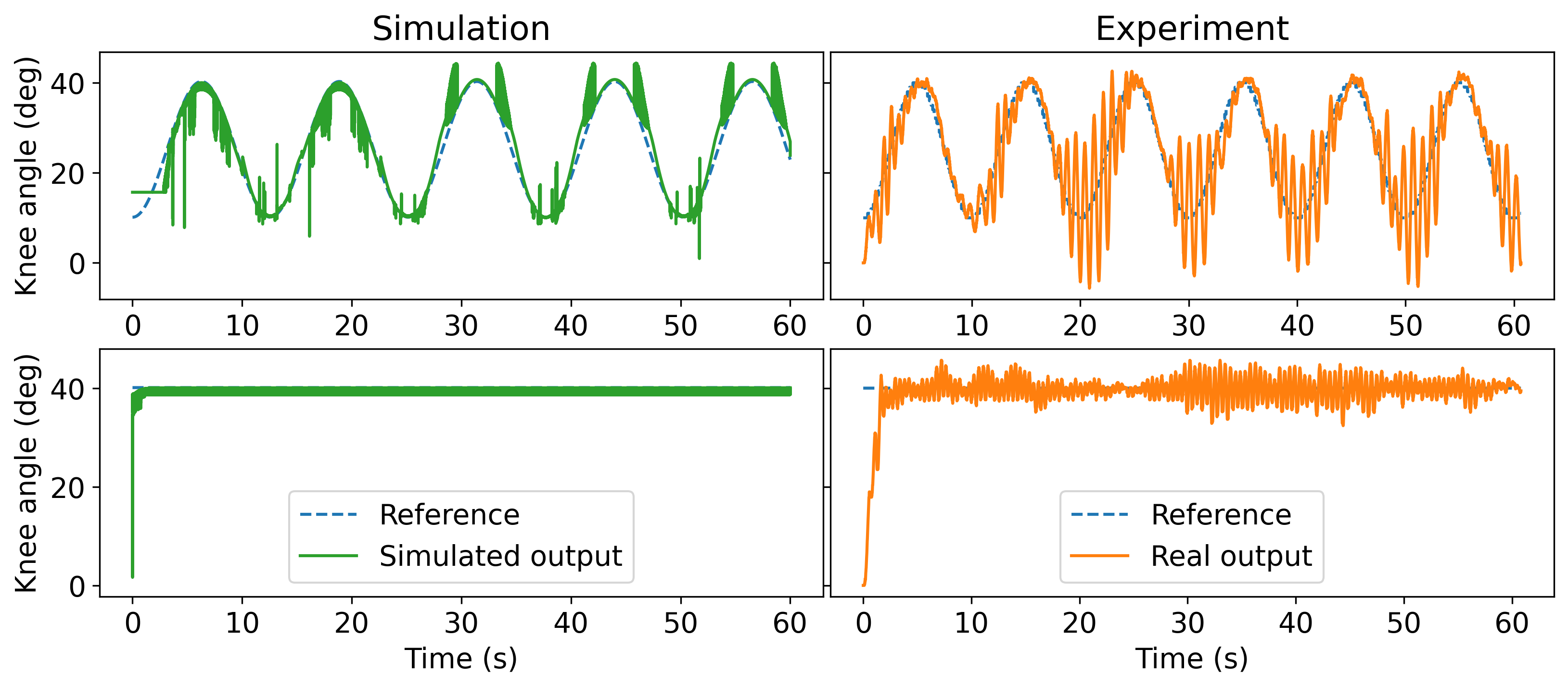}
	\caption{Comparison of simulation and real experiments for individual H3 using past rehabilitation data to identify the nonlinear model.}
	\label{fig:H3_simVSreal_emp}
\end{figure}

\begin{figure}[!h]
	\centering
	\includegraphics[width=1\linewidth]{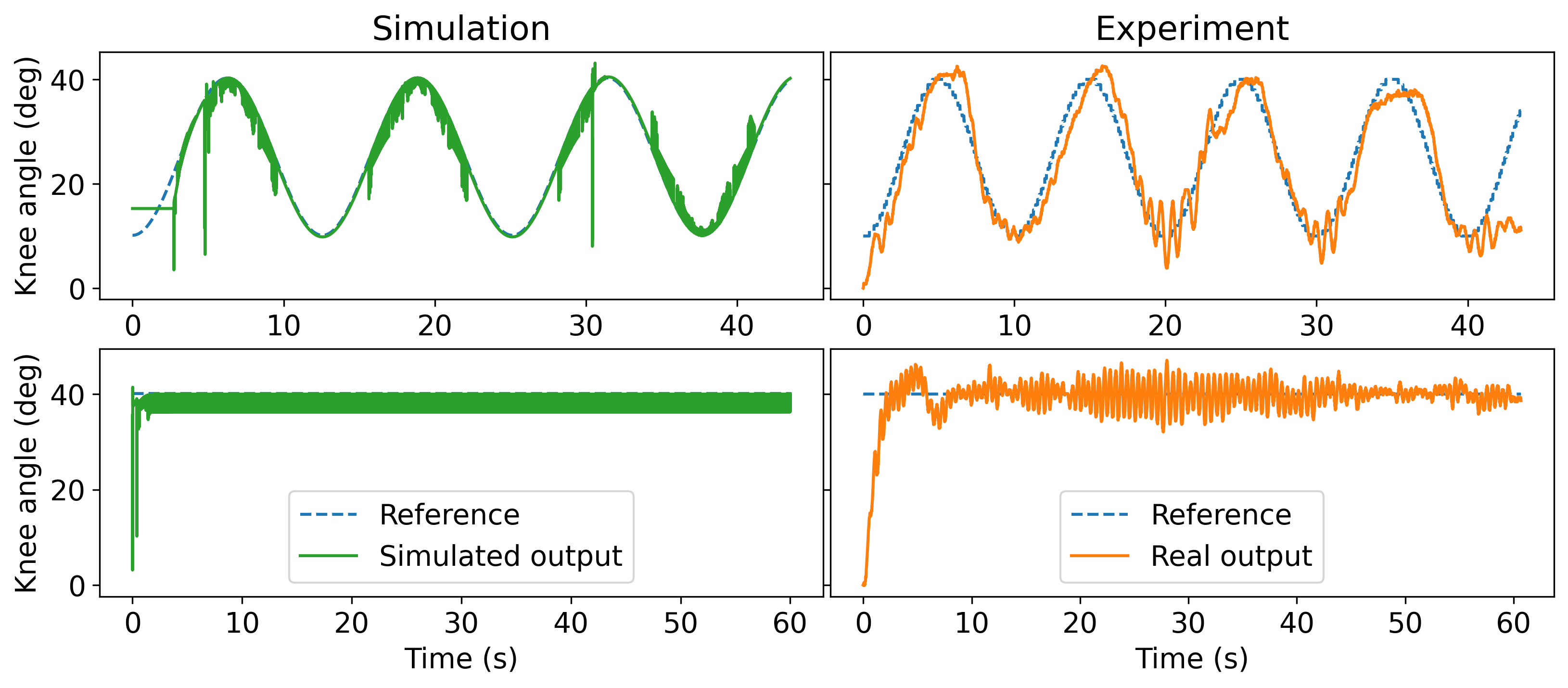}
	\caption{Comparison of simulation and real experiments for individual H6 using past rehabilitation data to identify the nonlinear model.}
	\label{fig:H6_simVSreal_emp}
\end{figure}

\begin{figure}[!h]
	\centering
	\includegraphics[width=1\linewidth]{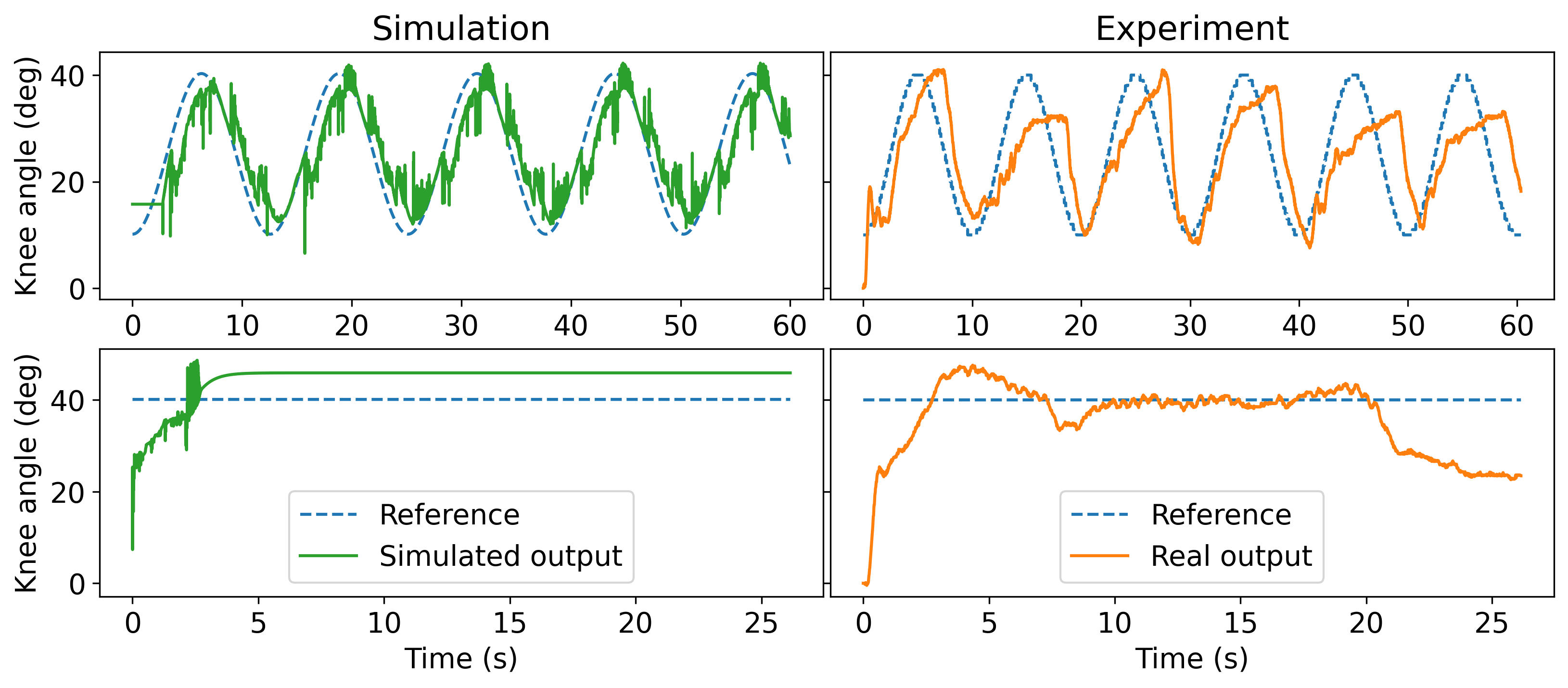}
	\caption{Comparison of simulation and real experiments for individual H7 using past rehabilitation data to identify the nonlinear model.}
	\label{fig:H7_simVSreal}
\end{figure}

\textcolor{black}{Table~\ref{tbl_gains} exhibits the fine-tuned gains (\(\alpha_1\); \(\alpha_2\); \(ks\); \(\beta\)) used for each RISE-based control-stimulation session and both trajectories.} Moreover, Figs.~\ref{fig:h1_results}-\ref{fig:h4_results} illustrate tracking results on both trajectories and their respective delivered PWs (Deliv. PWs) for individuals H1 (session \textit{v}), H2 (session \textit{ii}), and H4 (session \textit{v}). Figs.~\ref{fig:H2_simVSreal}-\ref{fig:H7_simVSreal} compare the results of simulation and real experiments based on empirical tuning or fine-tuned IGA gains. Figs.~\ref{fig:H2_simVSreal}-\ref{fig:H7_simVSreal} were selected for illustration purposes only, as the objective here is to highlight the benefits of using past data for the nonlinear system identification step.

\begin{table}[!h]
    \centering
    \scriptsize
    \caption{\textcolor{black}{RISE controller gains fine-tuned with IGA used in the experiments with the proposed methodology for both sine and step waves.}}
    \label{tbl_gains}
    \begin{tabular}{c c c c c c}
    \toprule
    \multirow{2}{*}{\textbf{Individual}} &\multirow{2}{*}{\textbf{Session}} &\multicolumn{2}{c}{\textbf{RISE controller gains (\(\alpha_1\); \(\alpha_2\); \(ks\); \(\beta\))}}\\\cmidrule{3-4}
    &&\textbf{Sine} &\textbf{Step}\\
    \midrule
    P1 &i &2.61; 3.34; 48.94; 1.78 &2.72; 3.57; 47.12; 1.54 \\\hline
    P2 &i &2.22; 3.54; 39.50; 1.40 &3.01; 1.91; 48.34; 2.65\\\hline
    \multirow{5}{*}{H1} &i  &3.23; 1.08; 24.74; 5.50 &1.37; 1.63; 54.03; 2.36\\
    &ii  &1.76; 2.28; 32.30; 2.39 &0.64; 1.66; 52.26; 4.00\\
    &iii  &3.23; 2.52; 27.33; 2.29 &2.30; 4.24; 59.26; 3.49 \\
    &iv  &2.40; 4.10; 27.05; 2.18 &3.12; 5.80; 43.162; 1.35\\
    &v  &3.07; 4.37; 21.73; 1.56 &2.61; 3.54; 39.50; 1.30\\\hline
    \multirow{5}{*}{H2} &i  &1.90; 3.50; 48.00; 3.00 &1.90; 3.50; 48.00; 3.00 \\
    &ii  &1.57; 2.37; 48.45; 1.05 &1.38; 1.34; 64.41; 3.72\\
    &iii   &1.47; 3.31; 30.01; 1.87 &3.54; 3.83; 54.88; 1.92 \\
    &iv &1.42; 3.68; 35.09; 1.99  &2.07; 1.75; 36.32; 1.96\\
    &v &2.03; 3.02; 36.07; 2.23  &2.17; 1.76; 38.53; 1.85\\\hline
    \multirow{5}{*}{H3} &i  &1.40; 2.50; 60.00; 3.40 &1.47; 2.63; 57.83; 3.36\\
    &ii  &2.12; 2.28; 73.74; 1.55 &2.77; 3.03; 57.17; 3.47\\
    &iii  &4.75; 4.01; 19.56; 2.73 &1.56; 3.95; 50.91; 3.05\\
    &iv &0.93; 2.69; 28.09; 2.45 &3.22; 3.99; 68.67; 1.26\\
    &v   &3.25; 3.45; 22.70; 3.23 &2.23; 2.85; 43.33; 2.03\\\hline
    \multirow{5}{*}{H4} &i   &1.92; 2.41; 69.71; 1.69  &1.92; 2.41; 69.71; 1.69\\
    &ii  &1.92; 4.14; 44.26; 1.50 &1.92; 2.41; 55.83; 1.69\\
    &iii  &3.85; 4.00; 21.51; 2.85 &1.22; 1.64; 30.44; 3.50\\
    &iv &1.63; 4.26; 23.74; 1.70 &1.03; 6.16; 66.38; 1.14\\
    &v  &2.24; 2.22; 28.37; 2.06 &2.12; 2.35; 46.93; 1.67\\\hline
    \multirow{3}{*}{H5} &i &3.36; 4.09; 53.19; 3.30 &3.65; 1.56; 76.66; 2.69\\
    &ii  &2.68; 6.85; 24.64; 3.13 &2.84; 1.51; 40.93; 2.54\\
    &iii  &3.21; 2.43; 51.30; 3.42 &1.16; 2.98; 45.15; 1.20\\\hline
    \multirow{2}{*}{H6} &i  &1.52; 2.50; 55.87; 1.67 &2.10; 1.08; 51.24; 1.93\\
    &ii &4.89; 4.89; 43.05; 2.36 &2.62; 5.22; 25.55; 3.65\\\hline
    \multirow{2}{*}{H7} &i   &3.72; 3.85; 45.16; 1.59 &2.75; 3.85; 68.51; 1.96\\
    &ii  &1.15; 5.96; 44.29; 1.20 &2.73; 5.79; 37.57; 2.44\\
    \bottomrule
    \end{tabular}
\end{table}

\end{document}